\documentclass[a4paper,11pt]{article}
\usepackage[a4paper,top=3.5cm,bottom=3cm,left=3cm,right=3cm]{geometry}

\usepackage[sectionbib]{natbib}
\bibpunct{[}{]}{;}{a}{}{,}

\usepackage[utf8]{inputenc} 
\usepackage[T1]{fontenc}    
\usepackage{hyperref}       
\usepackage{url}            
\usepackage{booktabs}       
\usepackage{amsfonts,amsmath}       
\usepackage{amsthm}
\usepackage{nicefrac}       
\usepackage{microtype}      
\usepackage{algorithm}
\usepackage{algpseudocode}
\usepackage{graphicx}
\usepackage{pdflscape}
\usepackage{longtable}
\usepackage{authblk}
\usepackage{array}
\usepackage{multirow}

\usepackage{latexsym}

\usepackage{color}

\def\colr{}
\def\colg{}

\def\colrr{}

\newcommand{\bs}{\boldsymbol}

\newcommand{\argmin}{\mathop{\rm argmin}\limits}

\newtheorem{assumption}{Assumption}

\usepackage{lineno}
\newcommand*\patchAmsMathEnvironmentForLineno[1]{
  \expandafter\let\csname old#1\expandafter\endcsname\csname #1\endcsname
  \expandafter\let\csname oldend#1\expandafter\endcsname\csname end#1\endcsname
  \renewenvironment{#1}
     {\linenomath\csname old#1\endcsname}
     {\csname oldend#1\endcsname\endlinenomath}}
\newcommand*\patchBothAmsMathEnvironmentsForLineno[1]{
  \patchAmsMathEnvironmentForLineno{#1}
  \patchAmsMathEnvironmentForLineno{#1*}}
\AtBeginDocument{
\patchBothAmsMathEnvironmentsForLineno{equation}
\patchBothAmsMathEnvironmentsForLineno{align}
\patchBothAmsMathEnvironmentsForLineno{flalign}
\patchBothAmsMathEnvironmentsForLineno{alignat}
\patchBothAmsMathEnvironmentsForLineno{gather}
\patchBothAmsMathEnvironmentsForLineno{multline}
}
\modulolinenumbers[5]

\title{Simultaneous Modeling of Disease Screening and Severity Prediction: A Multi-task and Sparse Regularization Approach}

\author[1]{Kazuharu Harada}
\author[2]{Shuichi Kawano}
\author[3]{Masataka Taguri}

\affil[1]{
    Tokyo Medical University, Japan \\
    Email: \texttt{haradak@tokyo-med.ac.jp}
}
\affil[2]{
    Kyushu University, Japan \\
    Email: \texttt{skawano@math.kyushu-u.ac.jp}
}
\affil[3]{
    Tokyo Medical University, Japan \\
    Email: \texttt{taguri@tokyo-med.ac.jp}
}
\date{\today}

\begin{document}
\maketitle

\begin{abstract}
{\colr Identifying clinically relevant biomarkers and developing predictive models are central challenges in biomedical research. Biomarkers are commonly used for disease screening, and some provide information not only on the presence or absence of a disease but also on its severity. Such biomarkers can contribute to treatment prioritization and support clinical decision-making. To address both disease screening and severity prediction, this paper focuses on regression modeling for ordinal outcomes with a hierarchical structure. When the response variable is a combination of the presence of disease and severity, such as \{{\it healthy, mild, intermediate, severe}\}, a straightforward approach is to apply the conventional ordinal regression model. However, such models may lack the flexibility needed to capture heterogeneity in how predictors relate to response levels, particularly when the response levels have a heterogeneous association structure with predictors. Therefore, this paper proposes a model that treats screening and severity prediction as separate tasks, along with an estimation method based on structural sparse regularization. This method is designed to leverage a shared structure between the tasks.} In numerical experiments, the proposed method demonstrated stable performance across many scenarios compared to existing ordinal regression methods.

\end{abstract}

\begin{center}
\fbox{
\parbox{0.9\textwidth}{
\small
This is the peer reviewed version of the following article: 
K. Harada, S. Kawano, and M. Taguri. Simultaneous Modeling of Disease Screening and Severity Prediction: A Multi-task and Sparse Regularization Approach. \textit{Expert Systems with Applications}. 2025, 
which has been published in final form at \url{https://doi.org/10.1016/j.eswa.2025.129408}. 
}}
\end{center}



\section{Introduction}
One of the important issues in biomedical research is the exploration of biomarkers and the construction of clinical prediction models. Biomarkers are physiological indicators or biomolecules used for various purposes such as disease screening, prognosis, and estimation of treatment effects, and they are widely used in clinical practice \citep{Biomarkers_Definitions_Working_Group2001-yd}. This paper focuses specifically on biomarkers used for disease screening, aiming to propose models and estimation methods for the exploration of biomarkers that are {\colr useful for screening and} can also predict the disease severity at screening. First, we will show examples of such {\colr multi-purpose} biomarkers.

Alpha-fetoprotein (AFP), Des-gamma-carboxyprothrombin, and Lens culinaris agglutinin-reactive fraction of AFP are well-known biomarkers for hepatocellular carcinoma. These biomarkers are measured through blood testing and serve as less invasive diagnostic markers for HCC. Furthermore, it has been reported that these markers are associated with prognosis and cancer stage \citep{Omata2017-gv, European_Association_for_the_Study_of_the_Liver_Electronic_address_easlofficeeaslofficeeu2018-ed}. Another example is neuroblastoma, which is a type of childhood cancer known to have a wide range of severity \citep{Irwin2021-jd}. It is known that neuroblastoma patients excrete specific metabolites in their urine, and conventionally, molecules such as homovanillic acid (HVA) and vanillylmandelic acid (VMA) have been used for disease screening. However, while these are highly effective in screening neuroblastoma, their association with severity is weak. Recently, a metabolomics study {\colr revealed that several markers, beyond HVA and VMA, are useful for screening and are associated with disease severity} \citep{Amano2024-ff}.

How should we take into account disease screening and severity prediction {\colr simultaneously}? A simple way would be {\colr to define the absence of disease as the lowest category and to set higher categories for increasing disease severity, treating it as an ordinal categorical variable.}
In regression analysis, an ordinal categorical variable is one of the most common types of response variables. A well-known regression model for ordinal responses is the cumulative logit model \citep[CLM, a.k.a. the proportional odds model;][]{McCullagh1980-en}. CLM assumes a continuous latent variable behind the response variable, and {\colr the level of the response variable} increases when the latent variable exceeds certain thresholds. CLM is also widely used in biomedical studies \citep[e.g.,][]{Muchie2016-al, Jani2017-zy}. However, this approach {\colr may have an issue when applying to our problem: CLM assumes} an identical relationship between the response and the predictor across all levels of the response variable, which is referred to as {\colr the} {\it parallelism assumption}. The parallelism assumption is questionable for the combined response because the two tasks, screening and severity prediction, do not necessarily share the same structure; in other words, the same predictors do not necessarily contribute in the same way to screening and severity prediction. One possible solution to this issue is to relax the parallelism assumption using a varying coefficient version of CLM, which we call non-parallel CLM (NPCLM). However, as discussed in Section 2, the varying coefficient approach could make the estimation and interpretation of the model difficult.

To address the problem of modeling the combined task, {\colr we propose a model based on the concept of multi-task learning. This approach aims to construct an effective prediction model and identify beneficial biomarkers, even in high-dimensional settings.} Multi-task learning is a machine learning approach where multiple models are trained simultaneously, leveraging shared information and structures across the tasks to improve predictive performance \citep{Caruana1997-lr, Argyriou2007-qn}. Multi-task learning is applied in many fields such as computer vision, natural language processing, and web applications \citep{Zhang2017-dz}. {\colr Furthermore, multi-task learning is increasingly being applied in biomedical research. For example, \citet{Wang2020-nw} introduced a multi-task learning framework to identify genes expressed across multiple cancer types; \citet{Moon2022-lo} developed a multi-task algorithm, MOMA, to integrate multi-omics data for accurate and interpretable disease classification across various clinical tasks; and more recently, \citet{Wu2024-hw} applied multi-task learning to predict cancer prognosis across types by leveraging shared patterns in RNA-seq and clinical data. As detailed in Section 4, our proposed method employs structured sparse regularization for parameter estimation across multiple hierarchical ordinal regression models. Multi-task learning with structured sparse regularization was initially discussed in studies such as \citet{Obozinski2006-mt}, \citet{Kim2010-lg}, and \citet{Zhou2010-bx}, and has continued to be explored in recent years in various directions, including applications to Gaussian copula models \citep{Goncalves2016-bx}, task clustering \citep{Okazaki2024-rz}, and extensions of the regularization terms themselves \citep[e.g.,][]{Fei2023-hd}.}
While there are not many examples of applying multi-task learning to ordinal regression, \citet{Xiao2023-hx}, for instance, {\colr deal with} multi-task learning for parallel ordinal regression problems.

{\colr 
The main contributions of this study are summarized in the following three points:
\begin{enumerate}
    \item {\bf Problem setting}: we formulate the joint task of screening and severity prediction, acknowledging that some biomarkers are linked to both disease presence and severity.
    \item {\bf Method}:  we propose a novel multi-task ordinal regression model, which incorporates structured sparse regularization to leverage shared structures between the tasks. Unlike CLM, our model offers greater flexibility, and in contrast to NPCLM, it remains valid and interpretable across the full predictor space.
    \item {\bf Findings and implications}: MtCLM shows strong performance in prediction and variable selection in both simulations and real data. The study highlights the potential of multi-task learning in biomedical research where related tasks share common structures.
\end{enumerate}
}

This paper is organized as follows. In Section 2, we review CLM, NPCLM, and related extensions. In Section 3, we propose a novel prediction model and also discuss its relationships with other categorical models and potential extensions. In Section 4, we introduce an estimation method using sparse regularization and an optimization algorithm. The proposed models employ structural sparse penalties to exploit the common structure between screening and severity prediction. 
{\colr In Section 5, we present the results of simulation experiments. We consider various structures between screening and severity prediction, and clarify under what conditions the proposed method performs well.}
In Section 6, we report the results of real data analysis. Finally, we provide concluding remarks in Section 7.


\section{Cumulative Logit Model}
For ordinal responses, CLM is one of the most popular regression models \citep{McCullagh1980-en,Agresti2010-at}. {\colr Let $Y_i\in\{0,1,\ldots,K\}:=\mathcal{Y}$ be the response and $X_i\in\mathcal{X}^p$ be the predictors for $i=1,\ldots,n$, where $\mathcal{X}^p$ is the $p$-dimensional predictor space, and the set $\mathcal{Y}$ is equipped with an ordinal relation corresponding to the natural numbers.} CLM is defined as
\begin{align*}
    \mathrm{logit}~\mathbb{P}(Y_i\le k \mid X_i) = \alpha_k + X_i^T{\bs\beta},~~~k=1,\ldots,K-1,
\end{align*}
where $\alpha_k\in\mathbb{R}$ and $\bs\beta\in\mathbb{R}^p$ are the intercepts and regression coefficients, and $\mathrm{logit}(u) = \log \{u / (1-u)\}~~(u\in(0,1))$. 
CLM has an important interpretation. Suppose that there is a latent continuous variable $Y^*\in\mathbb{R}$ behind the response and that $Y^*_i$ is associated with the predictors as
\begin{gather*}
    Y^*_i = X_i^T{\tilde{\bs\beta}} + \varepsilon_i,~~~~\varepsilon_i\sim Logistic(0,1),
\end{gather*}
where $\tilde{\bs\beta}\in\mathbb{R}^p$ represents the regression parameters for $Y^*_i$, and $Logistic(0,1)$ is the standard logistic distribution. Then, by defining $Y_i = k~\text{iff}~\tilde{\alpha}_{k-1}< Y^*_i \le \tilde{\alpha}_{k}$ with an increasing sequence $\{\tilde{\alpha}_k\}_{k=0}^K$ with $\tilde{\alpha}_0 = -\infty$ and $\tilde{\alpha}_K = \infty$, we have
\begin{gather*}
    \mathrm{logit}~\mathbb{P}(Y_i \le k\mid X_i) 
        = \mathrm{logit}~\mathbb{P}(Y_i^* \le \tilde{\alpha}_{k}\mid X_i) 
        = \mathrm{logit}~F_{\varepsilon}(\tilde{\alpha}_k - X_i^T{\tilde{\bs\beta}}) 
        = \tilde{\alpha}_k - X_i^T{\tilde{\bs\beta}},
\end{gather*}
where $F_{\varepsilon}$ is the cumulative distribution function of $Logistic(0,1)$.
By replacing $\alpha_k = \tilde{\alpha}_k$ for all $k\in\{1,\ldots,K-1\}$ and $\bs\beta = -\tilde{\bs\beta}$, we can see that this model is equivalent to CLM. Therefore, we can see $\alpha_k$ as the thresholds determining the class $Y_i$ based on the latent variable $Y_i^*$.
The linear functions $\alpha_k + X_i^T\bs\beta$ representing the log-odds of the cumulative probability of each level are all parallel because they share the same slope $\bs\beta$. The parallelism assumption is necessary to ensure that the model is valid in the sense that the conditional cumulative probability $\mathbb{P}(Y_i \le k \mid X_i = x)$ derived from CLM is monotonic at any given $x\in\mathcal{X}^p$ (e.g., \cite{Okuno2024-qt}).

The non-parallel CLM (NPCLM), sometimes called the non-proportional odds model, is the model with different slopes for each response level. This implies {\colr that} the conditional cumulative probability curves of NPCLM can be non-monotone. Such curves violate the appropriate ordering of the response probabilities. {\colr Thus,} NPCLM is more flexible and expressive than CLM, but it is valid on some restricted subspace of $\mathcal{X}^p$. Peterson et al. (1990) proposed a model that is intermediate between CLM and NPCLM \citep{Peterson1990-uu}. This partial proportional odds model is defined as follows:
\begin{gather}
    \mathrm{logit}~\mathbb{P}(Y_i\le k \mid X_i) = \alpha_k + X_i^T{\bs\beta} + X_i^T{\bs\gamma}_k,~~~k=1,\ldots,K-1,
\end{gather}
where $\bs\gamma_k$ is the {\colr level-specific} slope of the $k$th level.
This model is designed to capture the homogeneous effect {\colr of} $\bs\beta$ and the heterogeneous effect {\colr of} $\bs\gamma_k$. Still, similarly to NPCLM, this model is generally valid as a probability model only on the restricted subspace of $\mathcal{X}^p$.
To control the balance of flexibility and monotonicity, {\colr various regularization techniques} have been proposed. Wurm et al. (2021) use L1 and/or L2 {\colr norms on the} coefficient parameters $\bs\beta$ and $\bs\gamma_k$ of the partial proportional odds model \citep{Wurm2021-rv}. If the parallel assumption holds, then $\bs\gamma_k$ should be zero due to L1 penalization, and even when the parallel assumption is violated, the variability of $\bs\gamma_k$ is controlled by the penalties. Wurm et al. (2021) have also proposed an efficient coordinate descent algorithm \citep{Wurm2021-rv}, which is similar to that of the lasso regression \citep{Tibshirani1996-pc, Hastie2015-cn}. Tutz et al. (2016) have introduced the penalization on the difference of the adjacent regression coefficient \citep{Tutz2016-gl}. For NPCLM, their penalty term is $\sum_k\|\bs\gamma_{k+1} - \bs\gamma_{k}\|_2^2$, resulting in smoothed coefficients between the adjacent response levels. {\colr They also proposed to use the L1 penalty for the difference of the coefficients instead of the L2 penalty. Unlike the L2 penalty, the L1 penalty leads adjacent regression coefficients to be estimated as exactly equal when the parallelism assumption holds.} Additionally, they have proposed an algorithm based on the Alternating Direction Method of Multipliers (ADMM; \cite{Boyd2011-ak}). 

In our setting, the non-parallel models may be helpful, but as previously discussed, they can be non-monotone for given $x$. Furthermore, although these models are linear, they are sometimes not easy to interpret. If the estimated coefficients indicate that $\mathrm{logit}~\mathbb{P}(Y_i= healthy \mid X_i=x)$ and $\mathrm{logit}~\mathbb{P}(Y_i \le mild \mid X_i=x)$ have different coefficients with opposite sign for $X_i$, it may be difficult to understand, given that the event $\{Y_i \le mild, X_i=x\}$ includes $\{Y_i = healthy, X_i=x\}$. In the next section, we propose a multi-task learning approach that maintains monotonicity over the entire $\mathcal{X}^p$ and possesses a flexibility that allows it to capture the different structures between the tasks of screening and severity prediction.

\section{Proposed Model}

\subsection{Definition and interpretation}\label{sec31}
Let $Y_i\in\{0,1,\ldots,K\}$ be an ordinal outcome for which zero indicates {\colr case $i$ is healthy}, and $\{1,\ldots,K\}$ corresponds to disease severity. Our model, which we call {\it Multi-task Cumulative Linear Model (MtCLM)}, is defined as 
\begin{align}
    \mathrm{logit}~\mathbb{P}(Y_i = 0 \mid X_i) 
        =&~ \alpha + X_i^T{\bs\beta} \label{MtCLM1}\\
    \mathrm{logit}~\mathbb{P}(1 \le Y_i \le k \mid Y_i \ge 1, X_i) 
        =&~ \zeta_{k} + X_i^T{\bs\gamma},~~~k\in\{1,\ldots,K-1\} \label{MtCLM2}
\end{align}
where $\alpha, \zeta_{k}\in\mathbb{R}, {\bs\beta,\bs\gamma\in\mathbb{R}^p}$ are model parameters. We refer to model \eqref{MtCLM1} as the screening model and model \eqref{MtCLM2} as the severity model. The screening model is a simple logistic regression model, while the severity model {\colr is a} CLM for severity within the patient group. Note that the parameters are assumed to be variationally independent, meaning {\colr that} an estimator that maximizes the joint likelihood of \eqref{MtCLM1} and \eqref{MtCLM2} is equivalent to an estimator that maximizes the likelihood of \eqref{MtCLM1} and \eqref{MtCLM2} separately. When estimated using the penalized maximum likelihood method introduced in the next section, the proposed model can exploit the shared structure between screening and severity prediction.

Similarly to the CLM, the proposed model has a latent-variable interpretation. Let $Y^*_i, Y^{**}_i\in\mathbb{R}$ be the latent random variables, {\colr defined} as follows:
\begin{align*}
    Y^*_i =&~ X_i^T\tilde{\bs\beta} + \varepsilon^{*}_i,~~~~\varepsilon_i^*\sim Logistic(0,1), \\
    Y^{**}_i =&~ X_i^T\tilde{\bs\gamma} + \varepsilon^{**}_i,~~~~\varepsilon_i^{**}\sim Logistic(0,1), \\
    Y_i =&~ 0~\text{iff}~-\infty < Y^*_i \le \tilde{\alpha}, \\
    Y_i =&~ k~\text{iff}~ Y^*_i > \tilde{\alpha}~\mathrm{and}~\tilde{\zeta}_{k-1}< Y^{**}_i \le \tilde{\zeta}_{k}~~~\text{for}~k=1,\ldots,K,
\end{align*}
where $\tilde{\alpha}\in\mathbb{R}$ is a thresholding parameter, 
$\{\tilde{\zeta}_{k}\}_{k=0}^{K}$ is an increasing sequence {\colr such that} $\tilde{\zeta}_0 = -\infty$ and $\tilde{\zeta}_K = \infty$, $\tilde{\bs\beta},\tilde{\bs\gamma}\in\mathbb{R}^p$ are regression coefficients, and $(\varepsilon_i^*,\varepsilon_i^{**})$ are independent errors drawn from the standard logistic distribution. Then, we obtain
\begin{align*}
    \mathrm{logit}~\mathbb{P}(Y_i=0 \mid X_i) 
        =&~ \mathrm{logit}~\mathbb{P}(Y_i^*\le \tilde{\alpha}\mid X_i) = \tilde{\alpha} - X_i^T{\tilde{\bs\beta}}, \\
    \mathrm{logit}~\mathbb{P}(1\le Y_i \le k \mid Y_i\ge 1, X_i) 
        =&~ \mathrm{logit}~\mathbb{P}(Y_i^{**}\le {\tilde{\zeta}_k} \mid Y_i\ge 1, X_i) = \tilde{\zeta}_k - X_i^T\tilde{\bs\gamma}.
\end{align*}
That is, the proposed model {\colr assumes there are} latent variables $(Y^*_i, Y^{**}_i)$ behind screening and severity prediction and that they share the association structure {\colr with} $X_i$. As noted above, we leverage the shared structure between screening and severity prediction by penalized likelihood-based estimation.

{\colr It should be noted that MtCLM is not limited to its current form---specifically, the combination of screening and severity prediction.} Indeed, it could be reworked into a more comprehensive framework. {\colr For instance, the first-level model is not limited to binary screening, and a deeper hierarchy can be set.} More flexible models, such as multinomial logit models, can also be incorporated. Despite the potential generalizability, we choose to use the current form of MtCLM for two reasons. Firstly, a more generalized version could potentially increase complexity, making interpretation and practical application more challenging. Secondly, {\colr this paper aims to propose a method that captures the shared structure of the predictors in screening and severity prediction, and thus, more flexible models are beyond our scope.}

\subsection{Relationships to other categorical and ordinal models}\label{sec32}
{\colr Our model is related to other ordinal or non-ordinal categorical regression models.}
Let $g_0(X_i) = \alpha + X_i^T{\bs\beta}$ and $g_k(X_i) = \zeta_{k} + X_i^T{\bs\gamma}$. Then, the conditional cumulative probability for $Y$ is {\colr given by}
\begin{align*}
    &~ \mathrm{logit}~\mathbb{P}(Y_i\le k \mid X_i)\\
        =&~ \mathrm{logit}~\left\{\mathbb{P}(1 \le Y_i\le k \mid Y_i\ge 1, X_i)\mathbb{P}(Y_i \ge 1 \mid X_i) + \mathbb{P}(Y_i = 0 \mid X_i)\right\} \\
        =&~ \mathrm{logit}~\left[
            \sigma\{g_k(X_i)\}(1-\sigma\{g_0(X_i)\})  + \sigma\{g_0(X_i)\}
        \right],
\end{align*}
where $\sigma$ is a sigmoid function; i.e., the inverse of {\colr the} $\mathrm{logit}$ function.
We can see if $\sigma\{g_0(X_i)\} \rightarrow 0$, which means $\mathbb{P}(Y_i = 0\mid X_i)\rightarrow 0$, then the MtCLM reduces to CLM for $Y_i\in\{1,\ldots,K\}$. Also, we can see that MtCLM is non-parallel even if $\bs\beta = \bs\gamma$ because the gradient with respect to $X_i$ depends on $k$. The relationship between CLM and MtCLM is not straightforward due to the non-linearity of the sigmoid and logit functions.

We can see that the multinomial logit model (MLM) has a {\colr clearer} relationship with MtCLM than CLM. Let $\mathrm{logit}~\mathbb{P}(Y_i=k \mid X_i) = g^{(k)}_{MLM}(X_i)~(k=0,\ldots,K-1)$, where $g^{(k)}_{MLM}$ is the linear predictor for the $k$th level. Then, we obtain the following expression {\colr for} the conditional probability of MtCLM:
\begin{align}
    \mathrm{logit}~\mathbb{P}(Y_i=0 \mid X_i) =&~ g^{(0)}_{MLM}(X_i), \\
    \mathrm{logit}~\mathbb{P}(Y_i\le k \mid Y_i\ge1, X_i) 
        =&~ \log\frac{\mathbb{P}(1 \le Y_i\le k \mid X_i)}{\mathbb{P}(Y_i > k \mid X_i)} = \mathrm{log}~\frac{\sum_{j=1}^k \sigma\{g^{(k)}_{MLM}(X_i)\}}{1 - \sum_{j=0}^k \sigma\{g^{(k)}_{MLM}(X_i)\}}.
\end{align}

There are many other regression models for an ordinal response \citep{Agresti2010-at}.
The continuation ratio logit model assumes that the conditional probability $\mathbb{P}(Y_i = k+1 \mid Y_i \ge k, X_i)$ is logit-linear for all $k$, meaning that MtCLM is locally equivalent to the continuation ratio logit model at $k=1$. The adjacent category logit model expresses the logarithm of $\mathbb{P}(Y_i = k + 1 \mid X_i) / \mathbb{P}(Y_i = k \mid X_i)$ as a linear function. These models are flexible and do not violate the monotonicity of the cumulative probabilities. They are also useful and may be a good choice when there is a need for more flexible models. Specifically, if it is assumed that the variables important for lower and higher levels of severity prediction differ, these models could be quite useful. 
The partitioned conditional model \citep[PCM;][]{Zhang2012-so} is a regression model for a broad class of categorical responses with hierarchical structures. It {\colr was} introduced to model {\colr a} partially ordered response, where some of {\colr its} categories have ordinal relationships. This model encompasses MLM, CLM, and MtCLM as special cases. If more complex structures need to be modeled, it might be beneficial to refer to more general models like the PCMs. However, it is important to note that as the hierarchical structure becomes more complex, the devices for capturing shared structures among tasks can become more intricate.

As we see in Section 6, there is sometimes an insufficient number of cases at some levels of severity prediction. One reason {\colr why} MtCLM imposes a monotonicity assumption on the severity prediction model is to {\colr enable} stable estimation even in such situations. {\colr Furthermore}, the interpretation of the odds ratios varies model by model: the continuation ratio logit model estimates, for example, the odds ratio of {\it intermediate} to more than {\it intermediate} among the patients with {\it intermediate} or {\it severe}. The adjacent category logit model does not take into account the {\it severe} patients or {\it healthy} individuals when discussing {\it mild} vs. {\it intermediate} since it estimates the odds ratios of pairwise comparisons for adjacent categories. In contrast, MtCLM can be interpreted in the same way as a conventional logistic regression model for the screening model and as a conventional CLM for the severity prediction model. Although the final choice of model depends on the research purpose, MtCLM's strength also lies in its small {\colr difference} from the popular models.

\section{Sparse Estimation and Algorithm}
As noted in Section 3, we cannot exploit the shared structure of the two tasks, screening and severity prediction, through simple maximum likelihood estimation. Therefore, we employ a penalized maximum likelihood approach {\colr using structural sparse regularization}. 

\subsection{Penalized Likelihood for MtCLM}
The log-likelihood of MtCLM is given by
\begin{align}
    \ell(\mathbf{Y},\mathbf{X}, \alpha, {\bs \beta}, {\bs\zeta}, {\bs \gamma}) 
        =&~\log\mathcal{L}_1(\mathbf{Y},\mathbf{X}, \alpha, {\bs \beta}) + \log\mathcal{L}_2(\mathbf{Y},\mathbf{X}, {\bs\zeta}, {\bs \gamma}).
\end{align} Here, 
\begin{align}
    \mathcal{L}_1(\mathbf{Y},\mathbf{X}, \alpha, {\bs \beta})
        =&~ \prod_{i=1}^n \sigma(\alpha + X_i^T{\bs\beta})^{\mathbf{1}(Y_i= 0)}\left\{1 - \sigma(\alpha + X_i^T{\bs\beta})\right\}^{\mathbf{1}(Y_i\ge 1)}, \\
    \mathcal{L}_2(\mathbf{Y},\mathbf{X}, \bs\zeta, \bs\gamma)
        =&~ \prod_{i=1}^n\prod_{k=1}^K \left\{\sigma(\zeta_{k} + X_i^T\bs\gamma) - \sigma(\zeta_{k-1} + X_i^T\bs\gamma)\right\}^{\mathbf{1}(Y_i = k)},
\end{align}
where ${\bs\zeta} = (\zeta_{1},\ldots,\zeta_{K-1})^T$, $\mathbf{Y} = (Y_1,\ldots,Y_n)^T$, and $\mathbf{X} = (X_1,\ldots,X_n)^T$. Since we let $\zeta_{0}=-\infty$ and $\zeta_{K}=\infty$, $\sigma(\zeta_{0} + X_i^T\bs\gamma) = 0$ and $\sigma(\zeta_{K} + X_i^T\bs\gamma) = 1$ for all $X_i$. The log-likelihood for MtCLM is a sum of the log-likelihoods of the logistic regression model for screening and the CLM for severity prediction.

If the two {\colr probabilities}, $\mathbb{P}(Y_i=0 \mid X_i)$ and $\mathbb{P}(1\le Y_i\le k \mid Y_i\ge 1, X_i)$ have similar associations with the feature $X_j$, the corresponding regression coefficients $\beta_j$ and $\gamma_j$ are {\colr expected} to be similar. We implement this intuition using two types of structural sparse penalty terms, namely, the fused lasso-type penalty \citep{Tibshirani2005-rx} and the group lasso-type penalty \citep{Yuan2006-fw}.  
These two types of penalty terms are defined as follows:
\begin{gather}
    \mathcal{P}_{F}(\bs\beta,\bs\gamma; \lambda_F) = \lambda_F\sum_{j=1}^p|\beta_j - \gamma_j|,\\
    \mathcal{P}_{G}(\bs\beta,\bs\gamma; \lambda_G) = \lambda_G\sum_{j=1}^p\sqrt{\beta_j^2 + \gamma^2_j}
\end{gather}
where $\lambda_F,\lambda_G \ge 0$ are tuning parameters determining the intensity of regularization.

The estimator of MtCLM is defined as the minimizer of the penalized log-likelihoods:
\begin{gather}
    \min_{\alpha, {\bs \beta}, {\bs\zeta}, {\bs \gamma}} -\frac{1}{n}\ell(\mathbf{Y},\mathbf{X}, \alpha, {\bs \beta}, {\bs\zeta}, {\bs \gamma}) 
    + \mathcal{P}_{F}(\bs\beta,\bs\gamma; \lambda_F)
    + \lambda_{11}\|\bs\beta\|_1 + \lambda_{12}\|\bs\gamma\|_1, \label{eq:optimF}\\
    \min_{\alpha, {\bs \beta}, {\bs\zeta}, {\bs \gamma}} -\frac{1}{n}\ell(\mathbf{Y},\mathbf{X}, \alpha, {\bs \beta}, {\bs\zeta}, {\bs \gamma}) 
    + \mathcal{P}_{G}(\bs\beta,\bs\gamma; \lambda_G)
    + \lambda_{11}\|\bs\beta\|_1 + \lambda_{12}\|\bs\gamma\|_1, \label{eq:optimG}
\end{gather}
where $\lambda_{11},\lambda_{12} \ge 0$ are tuning parameters for L1 penalties {\colr on} $\bs\beta$ and $\bs\gamma$. We can enhance {\colr the} sparsity of $\bs\beta$ and $\bs\gamma$ by setting $\lambda_{11},\lambda_{12}$ to be nonzero. It would be possible to incorporate both $\mathcal{P}_{F}$ and $\mathcal{P}_{G}$ simultaneously, but we discuss them separately because they exploit the shared structure of the screening and severity prediction models in different ways. 

{\colr 
The fused lasso-type penalty imposes an L1 penalty on the difference between the two regression coefficients associated with covariate $X_j$. When it is expected that $\beta_j \approx \gamma_j$, as assumed above, this penalty encourages the estimated coefficients to reflect such similarity. In contrast, the group lasso-type penalty shrinks both $\beta_j$ and $\gamma_j$ exactly to zero when $X_j$ is weakly associated with both response variables---that is, when it contributes little to either screening or severity prediction. Conversely, if $X_j$ is relevant to at least one of the tasks (screening or severity prediction), the group lasso-type penalty tends to estimate both $\beta_j$ and $\gamma_j$ as non-zero, under the assumption that the variable contributes to both tasks. Compared to the latter, the former imposes a stronger constraint by encouraging the coefficient values themselves to be equal, thereby more aggressively leveraging structural similarity between tasks.
}

\begin{figure}
    \centering
    \includegraphics[width = 0.9\textwidth]{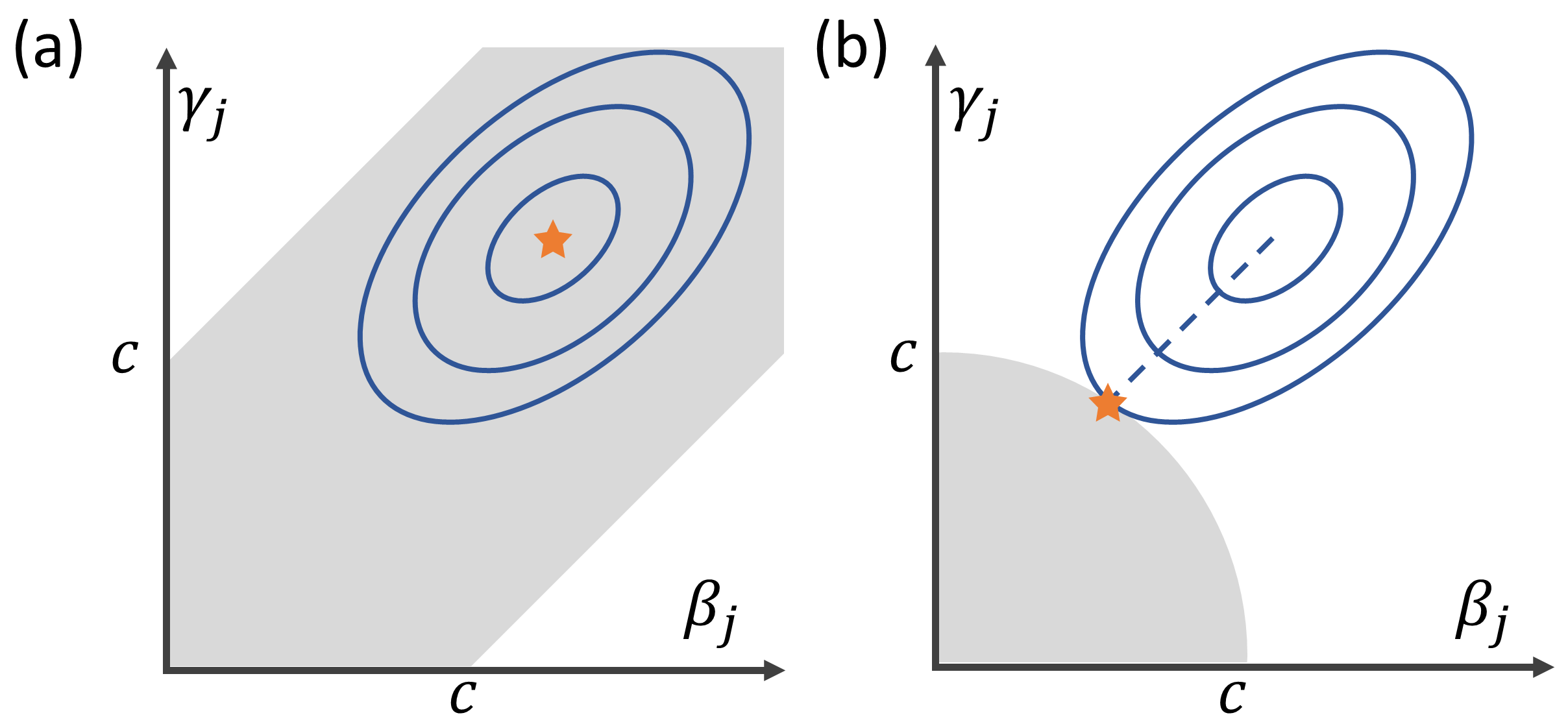}
    \caption{Illustrations for the structural sparse penalties and the optimal solution regarding the regression coefficients of $X_j$.
    Figure 1(a) shows the inequality constraint of the fused lasso type penalty (gray shading; $|\beta_j - \gamma_j|\le c$) and the log-likelihood function (blue contour lines), while Figure 1(b) shows the inequality constraint of group lasso type penalty (gray shading; $\sqrt{\beta_j^2 + \gamma^2_j}\le c$) and the log-likelihood function. {\colr The star marks indicate the optimal solutions.} When the unconstrained optimal solutions (center of the contour lines) satisfy $\beta_j\approx\gamma_j$ as {\colr in these} figures, it is understood that the fused lasso estimate is not biased, and the group lasso estimate is shrunk to the origin.}
    \label{fig:sparse1}
\end{figure}

Sparse regularization problems can be viewed as optimization problems under inequality constraints \citep{Hastie2015-cn}. Figure \ref{fig:sparse1} illustrates the optimization problems with inequality constraints for {\colr both the fused lasso and group lasso types of regularization}, and it explains how the solutions to these problems would be obtained. The group lasso-type regularization uniformly selects $X_j$ when it is related to $Y$, regardless of the {\colr position} of the true point $(\beta_j, \gamma_j)$, and a uniform bias toward the origin is introduced. On the other hand, the fused lasso type {\colr regularization} has a small bias when $\beta_j \approx \gamma_j$ is true, and in other cases, a bias enters according to the position of $(\beta_j, \gamma_j)$. 
For illustrations of cases {\colr where} $\beta_j \not\approx \gamma_j$, see \ref{app:Sparsity}.

Since the log-likelihood of the screening and severity prediction models and the penalty terms are convex \citep{Burridge1981-mn, Pratt1981-oa, Agresti2015-ez}, these penalized likelihood functions are also convex. The penalized log-likelihood functions have tuning parameters, and they can be selected by standard prediction-based techniques such as cross-validation {\colr (CV)}. In Section 5, we demonstrate parameter tuning of the proposed method {\colr using K-fold CV} based on the log-likelihood.
{\colr
Specifically, $\mathbf{Y}_{k}$ and $\mathbf{X}_{k}$ denote the data in the $k$-th fold of the K-fold partition, and $n_k$ is the sample size of that fold. Let $\hat{\alpha}^{\bs\lambda}_{-k}$, for example, denote the parameter estimates obtained from the data excluding the $k$-th fold at $\bs\lambda = (\lambda_{11},\lambda_{12},\lambda_F,\lambda_G)^T$. Then, the CV error to be minimized is defined as follows:
\begin{gather*}
    \text{CV}(\bs\lambda) = -\frac{1}{K}\sum_{k=1}^K \frac{1}{n_{k}}\ell(\mathbf{Y}_{k},\mathbf{X}_{k}, \hat{\alpha}^{\bs\lambda}_{-k}, \hat{\bs \beta}^{\bs\lambda}_{-k}, \hat{\bs\zeta}^{\bs\lambda}_{-k}, \hat{\bs \gamma}^{\bs\lambda}_{-k}).
\end{gather*}
}

\subsection{Alternating Direction Method of Multipliers for MtCLM}
The Alternating Direction Method of Multipliers (ADMM) is a popular algorithm used for convex optimization problems where the objective function is the sum of two convex functions. ADMM introduces {\colr auxiliary parameters} and breaks down the problem into smaller, more tractable components. It is widely applicable to sparse estimation methods, including the fused lasso and the group lasso \citep{Boyd2011-ak}.

To derive the ADMM algorithm for MtCLM, we prepare another expression for the penalties. Let $\bs\Theta = (\bs\beta~\bs\gamma)$, and the penalty terms are re-expressed as 
\begin{align*}
    \mathcal{P}_{F}(\bs\Theta; \lambda_F) 
        =&~ \lambda_F \|{\bs\Theta}{\bs d}\|_1,\\
    \mathcal{P}_{G}(\bs\beta,\bs\gamma; \lambda_F) 
        =&~ \lambda_G\sum_{j=1}^p \|{\bs\theta}_j\|_2,
\end{align*}
where ${\bs d} = (1 -1)^T$, and ${\bs\theta}_j~(j=1,\ldots,p)$ are row vectors of $\bs\Theta$.
The fused-lasso penalty can be expressed as a special case of the generalized lasso \cite{Tibshirani2011-vu}, but we write it as above for convenience.

We derive the ADMM algorithm for the problem \eqref{eq:optimF}. The optimization problem \eqref{eq:optimF} is equivalent to the following one, which introduces redundant parameters ${\bs a}\in\mathbb{R}^{p}$ and ${\bf B} = ({\bs b}_{\cdot 1}~ {\bs b}_{\cdot 2})\in\mathbb{R}^{p\times 2}$:
\begin{gather}\label{eq:optimADMM_F}
    \begin{array}{c}\displaystyle
        \min_{\alpha,\bs\zeta,\bs\Theta, {\bs a}, {\bf B}}~-\frac{1}{n}\ell(\mathbf{Y},\mathbf{X}, \alpha,\bs\zeta,\bs\Theta,{\bf B}) + \lambda_{F}\|{\bs a}\|_1 
        + \lambda_{11}\|{\bs b}_{\cdot 1}\|_1 + \lambda_{12}\|{\bs b}_{\cdot 2}\|_1 \\
        \text{subject to} ~~~ \bs\Theta{\bs d} = {\bs a}, \bs\Theta = \mathbf{B}.
    \end{array}
\end{gather}
The augmented Lagrangian of this problem is defined as
\begin{align}
    L(\alpha,\bs\zeta,\bs\Theta, {\bs a},{\bf B})
        =&~ -\frac{1}{n}\ell(\mathbf{Y},\mathbf{X}, \alpha,\bs\zeta,\bs\Theta,{\bf B}) + \lambda_{F}\|{\bs a}\|_1 + {\bs u}^T(\bs\Theta{\bs d} - {\bs a})
            + \frac{\mu_F}{2}\|\bs\Theta{\bs d} - {\bs a}\|^2_2 \nonumber\\
        &~~~~ + \lambda_{11}\|{\bs b}_{\cdot 1}\|_1 + \lambda_{12}\|{\bs b}_{\cdot 2}\|_1
            + \mathrm{tr}\left\{\mathbf{V}^T(\bs\Theta - \mathbf{B})\right\}
            + \frac{\mu_1}{2}\|\bs\Theta - \mathbf{B}\|^2_F, \label{eq:auglag}
\end{align}
where $\bs u$ and $\mathbf{V}=({\bs v}_{\cdot 1}~{\bs v}_{\cdot 2})$ are Lagrange {\colr multipliers}, and $\{\mu_F, \mu_1\}$ are tuning parameters for optimization.
In ADMM, the parameters are updated in sequence to minimize \eqref{eq:auglag}. The Lagrange multipliers are updated by the gradient {\colr ascent}. Given the parameters of the previous step, the updating formulae are given below:
\begin{align}
    (\alpha^{t+1}, \bs\zeta^{t+1},\bs\Theta^{t+1})
        =&~ \argmin_{\bs\alpha,\bs\zeta,\bs\Theta}~
            L(\alpha,\bs\zeta,\bs\Theta, {\bs a}^{t}, \mathbf{B}^{t}), \label{eq:upd1}\\
    \bs{a}^{t+1}
        =&~ \argmin_{\bs{a}}~ \lambda_{F}\|{\bs a}\|_1
            + \bs{u}^{tT}(\bs{\Theta}^{t+1}{\bs d} - \bs{a}) 
            + \frac{\mu_F}{2}\|\bs{\Theta}^{t+1}{\bs d} - \bs{a}\|_2^2, \label{eq:upd2}\\
    \mathbf{B}^{t+1}
        =&~ \argmin_{\mathbf{B}}~ \lambda_{11}\|{\bs b}_{\cdot 1}\|_1 + \lambda_{12}\|{\bs b}_{\cdot 2}\|_1 \nonumber\\
        &~~~~~~~~~~~~ + \mathrm{tr}\left\{\mathbf{V}^{tT}(\bs\Theta^{t+1} - \mathbf{B})\right\}
        + \frac{\mu_1}{2}\|\bs\Theta^{t+1} - \mathbf{B}\|^2_2, \label{eq:upd3}\\
    \bs{u}^{t+1} 
        =&~ \bs{u}^{t} + {\colr\mu_F}(\bs{\Theta}^{t+1}{\bs d} - \bs{a}^{t+1}), \label{eq:upd4}\\
    \mathbf{V}^{t+1} 
        =&~ \mathbf{V}^{t} + {\colr\mu_1}(\bs{\Theta}^{t+1} - \mathbf{B}^{t+1}).\label{eq:upd5}
\end{align}
The small problem of \eqref{eq:upd1} does not have an explicit solution, so it must be solved using an iterative algorithm. Since the target function of \eqref{eq:upd1} is convex, it can be solved using an off-the-shelf solver. In our implementation, we used the {\it optim} package in R. The problems \eqref{eq:upd2} and \eqref{eq:upd3} have explicit solutions:
\begin{align}
    \bs{a}^{t+1} 
        =&~ S(\bs{\Theta}^{t+1}{\bs d} + \mu_F^{-1}\bs{u}^{t}, \mu_F^{-1}\lambda_F), \label{eq:upd2_ex}\\
    \mathbf{\bs b}_{\cdot 1}^{t+1} 
        =&~ S(\bs{\beta}^{t+1} + \mu_1^{-1}{\bs v}_{\cdot 1}^{t+1}, \mu_1^{-1}\lambda_{11}),\label{eq:upd3_ex1}\\ 
    \mathbf{\bs b}_{\cdot 2}^{t+1} 
        =&~ S(\bs{\gamma}^{t+1} + \mu_1^{-1}{\bs v}_{\cdot 2}^{t+1}, \mu_1^{-1}\lambda_{12}), \label{eq:upd3_ex2}
\end{align}
where the function $S(z,\xi) = \mathrm{sign}(z)(|z|-\xi)_+$ is the soft-thresholding operator, which is applied element-wise to a vector.
Repeat steps \eqref{eq:upd1} through \eqref{eq:upd5} until an appropriate convergence criterion is met to obtain the final estimate. {\colr 
The algorithm for \eqref{eq:optimG} is derived in a similar manner, as shown in \ref{app:ADMM1}. In addition, other subsections of \ref{app:ADMM} provide the gradient of the augmented Lagrangian, the pseudocode, and a brief discussion on the convergence of ADMM, supported by both theoretical and empirical analysis.
}

\section{Numerical Experiments}

In this section, we perform numerical experiments to compare the proposed and existing methods in prediction and variable selection performance in several scenarios. In all scenarios, the sample size is set to 300. Response $Y$ has four levels $\{0,1,2,3\}$, generated by thresholding the latent {\colr variables} $Y^*$ and $Y^{**}$. Predictors are generated from standard normal distributions that are independent of each other. The dimensions of the predictors are set to four values for each scenario: 75, 150, 300, and 450. As described in Table \ref{tab:scenarios}, five scenarios with different true structures, in which 10 to 20 predictors are truly relevant to the response, are set up to account for various situations.

\begin{table}[ht]
    \centering
    \caption{Scenarios for the numerical experiments}
    \begin{tabular}{ccp{10cm}l}\hline
        No. &  Name & Description \\\hline
        1 & Parallel & All levels of $Y$ follow the parallel CLM.\\ \vspace{1mm}
        2 & Identical & MtCLM. All relevant predictors have regression coefficients with the same signs in screening and severity prediction.\\\vspace{1mm}
        3 & Almost Inverse & MtCLM. The relevant predictors are common in both screening and severity prediction, but the signs of the regression coefficients are almost inverse for each task.\\\vspace{1mm}
        4 & Similar & MtCLM. Many of the relevant predictors are common between the screening and severity models, but some are different. \\\vspace{1mm}
        5 & Almost Independent & MtCLM. Only two predictors are shared between both tasks.\\ \hline
    \end{tabular}
    \label{tab:scenarios}
\end{table}
Details on the simulation models are found in \ref{app:experiments}. Note that Scenario 2 appears to satisfy the parallelism assumption, but {\colr not exactly as mentioned in} Section \ref{sec32}. Scenarios 2 to 5 do not meet the parallelism assumption, and in particular, Scenarios 3 and 5 are in serious violation.

The proposed and existing methods include the MtCLM with three types of penalties (L1, L1 + Group lasso, and L1 + Fused lasso), the L1-penalized logistic regression (\textit{glmnet} package, only for screening), and the L1-penalized parallel and non-parallel CLM (\textit{ordinalNet} package \citep{Wurm2021-rv}). All these methods have tuning parameters. For the proposed methods, $\lambda_{11}$ is chosen from $(0.01, 0.05)$, $\lambda_{12}$ is chosen from $(0.05, 0.1)$, and $\lambda_F, \lambda_G$ are chosen from $(0, 0.01, 0.05)$ using 5-fold CV {\colr as described in Section 4.1}. Similarly, the tuning parameters of the existing methods are chosen from $(0.01, 0.05, 0.1)$ based on {\colr the out-of-sample likelihood evaluated via} a 5-fold CV. {\colr The screening performance is evaluated using the Area Under the Receiver Operating Characteristic Curve (ROC-AUC) and the F1 score when classified with a threshold of 0.5, based on the prediction probabilities.}
{\colr The performance for the ordinal response, i.e., the joint task of screening and severity prediction across all categories (0, 1, 2, 3)}, is evaluated using Accuracy, Mean Absolute Error (MAE), and Kendall's $\tau$, which is a measure of concordance for ordinal categories. The performance in variable selection is evaluated in terms of the power and false discovery rate (FDR) based on the ground truth. These metrics are calculated for the three types of MtCLM {\colr based on the number of regression coefficients estimated to be non-zero and the number of true non-zero regression coefficients}, for scenarios where the true structure has a hierarchical structure of MtCLM. These measures are calculated over 30 simulations for each scenario. Details about the simulation settings and the evaluation measures are found in \ref{app:experiments}. 

\begin{figure}[htbp]
    \centering
    \includegraphics[width = \textwidth]{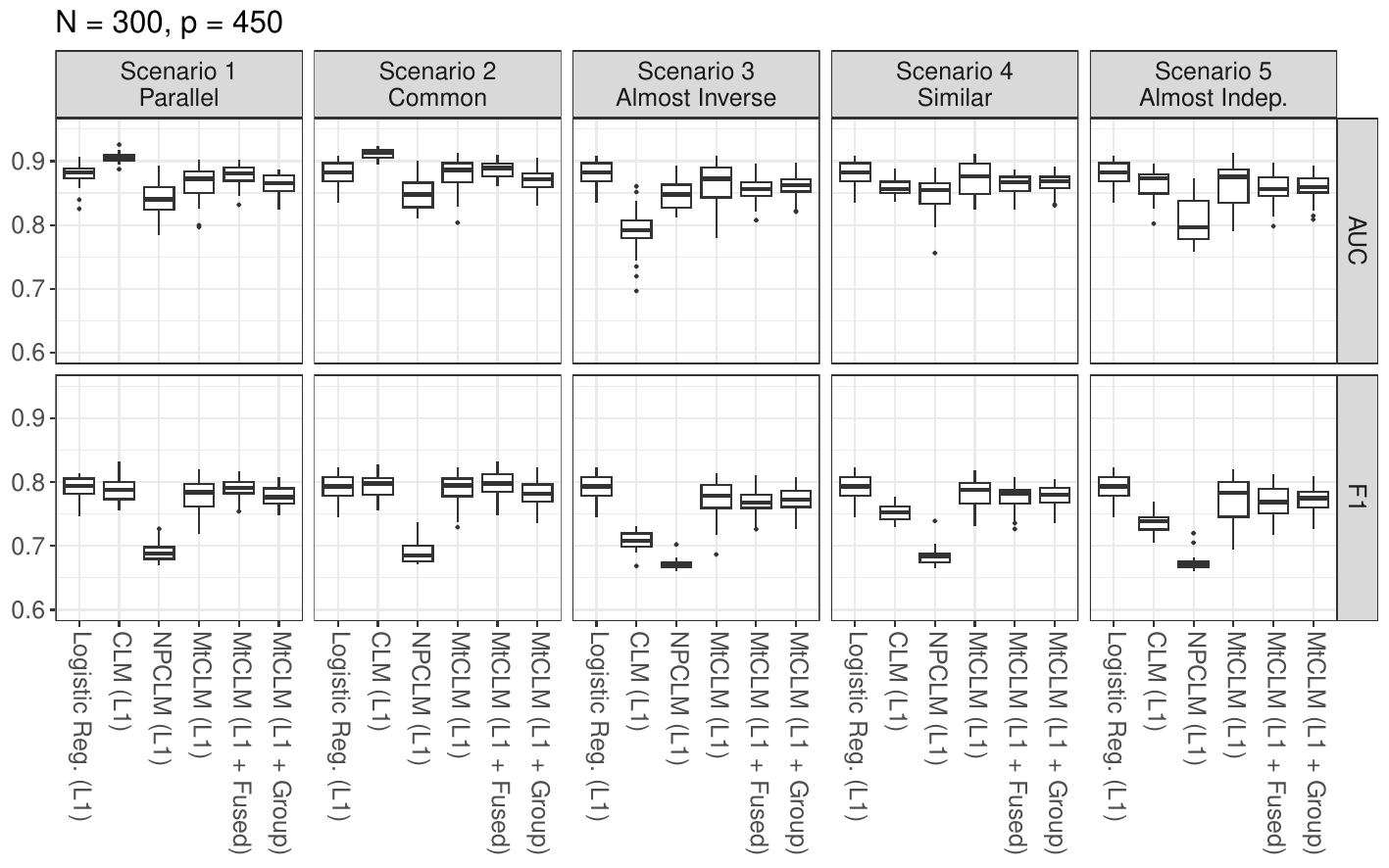}
    \caption{Comparison of the proposed and existing methods for the screening (0/1 classification) with 450-dimensional predictors.}
    \label{fig:performance_in_diag}
\end{figure}

Figure \ref{fig:performance_in_diag} shows the screening performance under the situation with high-dimensional predictors ($N=300, p=450$). 
The penalized logistic regression, which did not model the severity, showed similar performance across all scenarios and metrics. The penalized CLM exhibited superior performance, particularly in AUC, in Scenario 1, where the parallelism assumption holds, and Scenario 2, where the situation is close to Scenario 1. This result implies that, under appropriate conditions, considering severity can improve screening performance even when tackling a screening task. On the other hand, the NPCLM performed worse than the other methods in all scenarios. This performance degradation can also be attributed to the defect in non-parallel models described in Section 2. Comparing the three MtCLMs, there was no significant difference in screening performance regardless of the penalty used, but the MtCLM (L1) showed slightly larger variability in performance compared to the others. The MtCLM consistently demonstrated performance comparable to or slightly lower than the penalized logistic regression across all scenarios. By making the model more flexible compared to CLM, the performance remained stable even in scenarios where the two tasks had almost no similar structure. Furthermore, in the scenarios with high commonality between tasks (Scenarios 1, 2, and 4), it showed performance comparable to the penalized logistic regression, suggesting that it effectively leveraged the shared structure in appropriate situations despite increased variance due to increased parameters. In the settings with lower dimensional predictors, we obtained similar results; however, it is worth mentioning that the MtCLM slightly outperformed the penalized logistic regression in some scenarios when $p=75$ (See Figure \ref{fig:performance_in_diag_075} in \ref{app:results}).

\begin{figure}[htbp]
    \centering
    \includegraphics[width = \textwidth]{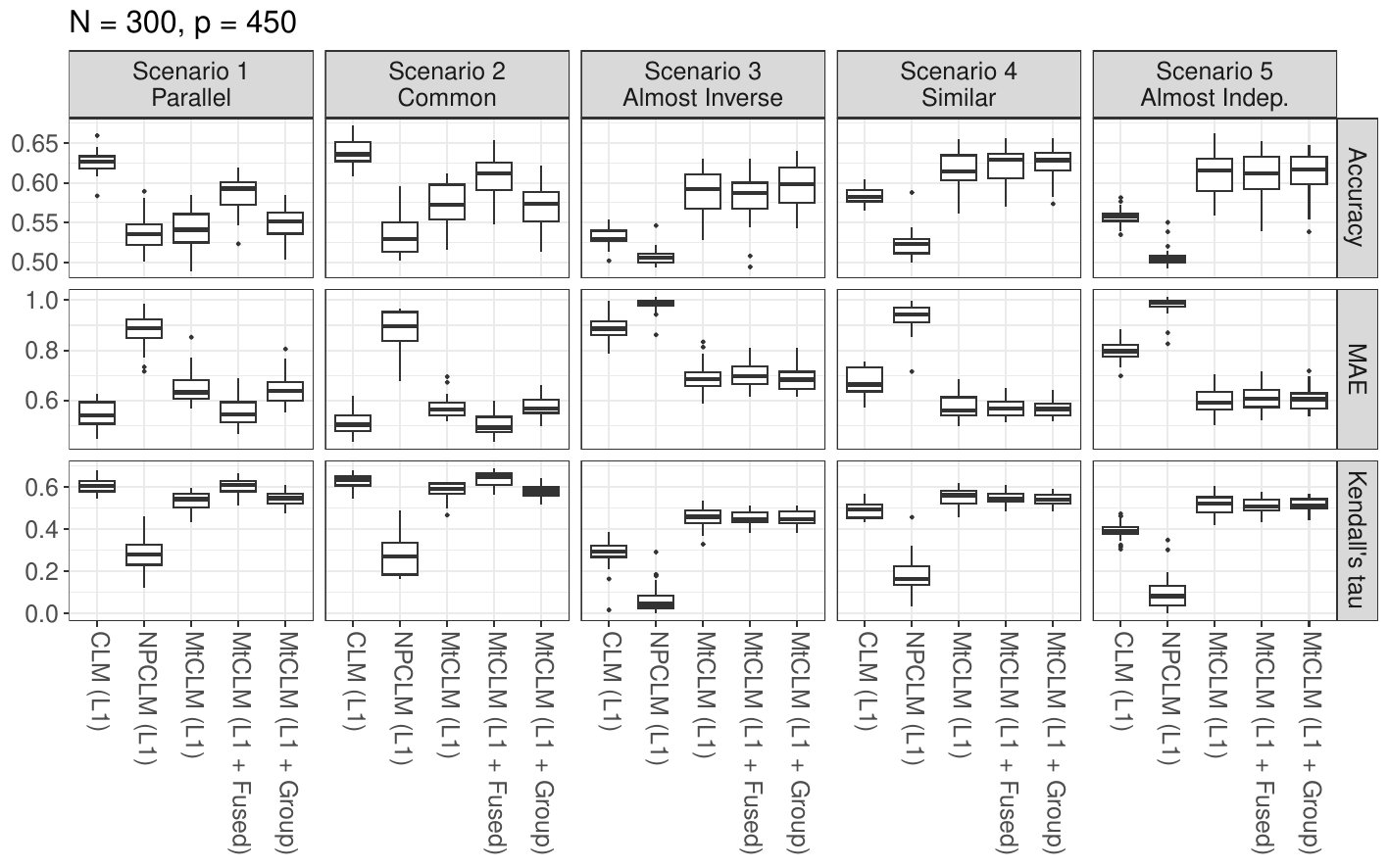}
    \caption{Comparison of the proposed and existing methods for the joint task of screening and severity prediction with 450-dimensional predictors.}
    \label{fig:performance_in_joint}
\end{figure}

Predictive Performance for the ordinal responses, namely, the joint task of screening and severity prediction is presented in Figure \ref{fig:performance_in_joint}.

As with screening, in Scenarios 1 and 2, where the structure of both tasks is highly similar, the penalized CLM demonstrated high performance. Particularly in terms of Accuracy, the CLM exceeded the others; however, in terms of concordance and MAE, MtCLM (L1 + Fused) was comparable. The penalized NPCLM failed to perform well in all scenarios, and the MtCLM consistently showed high performance across all scenarios, {\colr especially in Scenarios 1 and 2, the fused lasso penalty worked effectively}. The large variance of MtCLM (L1) was also similar to the screening performance results. {\colr In addition}, MtCLM (L1 + Group) slightly outperformed others in terms of Accuracy in scenarios 3, 4, and 5.
When the predictors are of lower dimensions ($p=75$), MtCLM (particularly L1 + Fused) demonstrated performance equal to or better than CLM in Scenarios 1 and 2. Additionally, in cases where the same variables are related to the response, though in different directions, as in Scenario 3, MtCLM (L1 + Group) performed relatively well.

\begin{figure}[htbp]
    \centering
    \includegraphics[width = \textwidth]{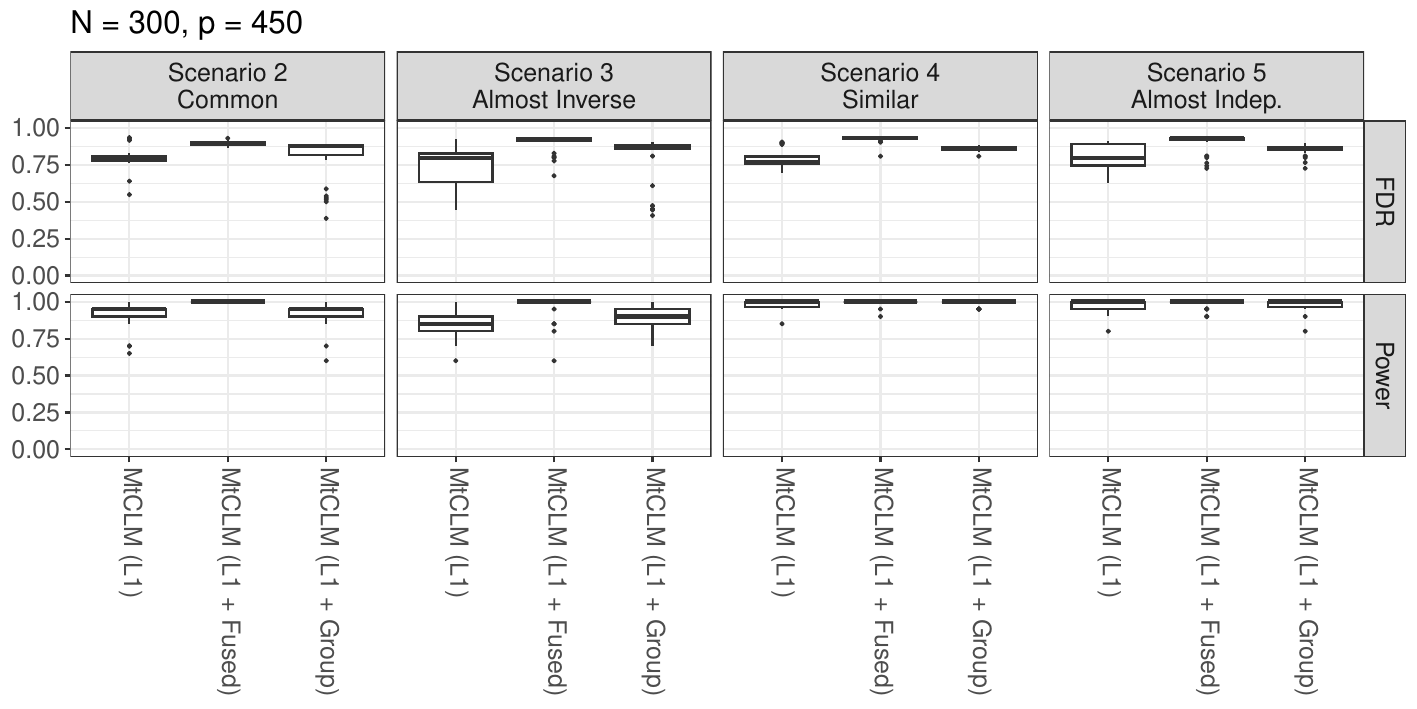}
    \caption{Comparison of the proposed methods in variable selection among 450-dimensional predictors.}
    \label{fig:performance_in_selection}
\end{figure}

In variable selection, the results were compared among the variants of MtCLM (Figure \ref{fig:performance_in_selection}). 
Overall, as the dimensionality of the predictors increased, the number of false discoveries also increased. In our settings, all methods were able to select most variables that are truly associated with the response. The FDR varied depending on the choice of penalty term. When using the fused lasso penalty, especially in scenarios 3, 4, and 5, the FDR was very high, and many variables unrelated to the response were selected. When using the L1 penalty alone, the FDR was lower in many scenarios on average, but the variability of the FDR was the highest. When using the group lasso penalty, in scenarios 2 and 3, it showed detection power similar to that of using the L1 penalty alone, and the FDR was relatively lower and more stable than with the fused lasso penalty.

In summary, when there is a hierarchical structure in ordinal responses, if the structures between {\colr the} hierarchies are highly similar, or if the parallelism assumption is believed to hold, CLM is highly effective. {\colr Moreover, even when the primary goal is classification at the first hierarchy---namely, screening---modeling the response as an ordinal variable that includes severity levels may improve screening performance.} 
{\colr However, given the possibility of structural differences between hierarchies, a model that accounts for such differences---such as MtCLM---should be employed. As demonstrated across the five scenarios, MtCLM performs effectively even when the hierarchical structures differ substantially, despite the use of penalties like the fused lasso that promote shared structures.}
Furthermore, when the structures are well-shared between hierarchies, the fused lasso penalty is particularly effective, performing comparably to CLM. The proposed methods are said to be adaptive to the data and robust to structural differences, as they improve performance by leveraging commonalities between hierarchies when structures are highly shared and {\colr fit each task separately when commonalities are poor.}

\section{Real Data Analysis}
\subsection{Pancreatic Ductal Adenocarcinoma Dataset}\label{app:PDAC}
In this section, we report the results of applying the proposed methods and some existing methods.

\cite{Debernardi2020-tv} provide an open and cleaned dataset on the concentration of certain proteins in the urine of healthy individuals and patients with pancreatic ductal adenocarcinoma (PDAC) or benign hepatobiliary diseases. The dataset was downloaded from Kaggle datasets on July 3rd, 2023 \footnote{https://www.kaggle.com/datasets/johnjdavisiv/urinary-biomarkers-for-pancreatic-cancer?resource=download}. The Debernardi dataset consists of 590 individuals and includes the {\colr concentrations} of five molecules (creatinine, LYVE1, REG1B, TFF1, REG1A), cancer diagnosis, cancer stage, and certain demographic factors. They reported that a combination of the three molecules, LYVE1, REG1B, and TFF1, showed good predictive performance in detecting PDAC. To combine the tasks of screening and severity prediction, we defined the response variable as summarized in Table \ref{tab:composite_res}. For further descriptive information on the dataset, refer to \ref{app:PDAC}. 

\begin{table}[ht]
    \centering
    \caption{Summary for the composite response.}
    \begin{tabular}{ccp{7cm}r}
        \hline
        Levels & Hierarchy & Definition & \# of Cases \\ \hline
        0 & 1st & No PDAC, including benign hepatobiliary diseases  & 391\\
        1 & 2nd & PDAC at stages I, IA, and IB & 16\\
        2 & 2nd & PDAC at stages II, IIA, and IIB & 86\\
        3 & 2nd & PDAC at stage III & 76\\
        4 & 2nd & PDAC at stage IV & 21 \\
        \hline
    \end{tabular}
    \label{tab:composite_res}
\end{table}

We applied the methods CLM ({\it polr}), CLM (L1), MtCLM (L1), MtCLM (L1 + Fused), and MtCLM (L1 + Group) to this composite response, using age and the {\colr concentrations} of molecules as predictors. Note that REG1A was excluded from the analysis due to the presence of missing values in about half of the cases. For the methods with tuning parameters, we selected them using 5-fold cross-validation. 

Table \ref{tab:Debernardi_coef} shows the estimated regression coefficients of these methods.
\begin{table}[ht]
    \centering
    \caption{Estimated regression coefficients for each model. CLM is a conventional ordinal regression model, and MtCLM is the proposed one. L1, Fused, and Group denote the types of regularization terms. The hyphen indicates that the regression coefficient is zero, in other words, the variable is not selected.}
    \begin{tabular}{lllrrrrr}
    \hline
        \multicolumn{1}{c}{Model} & \multicolumn{1}{c}{Regularization} & \multicolumn{1}{c}{Task} & \multicolumn{1}{c}{age} & \multicolumn{1}{c}{creatinine} & \multicolumn{1}{c}{LYVE1} & \multicolumn{1}{c}{REG1B} & \multicolumn{1}{c}{TFF1} \\
        \hline
        CLM & None & Overall & 0.634 & $-0.286$ & 2.381 & 0.474 & 0.023 \\
         & L1 & Overall & 0.594 & $-0.169$ & 1.847 & 0.471 & - \\
         &  &  &  &  &  &  &  \\
        MtCLM & L1 & Screening & 0.711 & $-0.467$ & 2.001 & 0.594 & 0.113 \\
         &  & Sev. Pred. & - & 0.259 & - & - & - \\
         & L1 + Fused & Screening & 0.677 & $-0.376$ & 1.761 & 0.582 & 0.066 \\
         &  & Sev. Pred. & - & 0.241 & - & - & - \\
         & L1 + Group & Screening & 0.680 & $-0.377$ & 1.666 & 0.592 & 0.089 \\
         &  & Sev. Pred. & - & 0.249 & - & - & - \\
         \hline
    \end{tabular}
    \label{tab:Debernardi_coef}
\end{table}
The coefficients of CLM ({\it polr}) and CLM (L1) indicated that age, LYVE1, and REG1B were positively associated with the composite response, and that creatinine was negatively associated. On the other hand, the coefficients of the MtCLMs suggested that creatinine was negatively associated with the presence of cancer but positively associated with cancer severity. These inverse relationships can also be observed {\colr in} the descriptive analysis (see Figure \ref{fig:Debernardi_univ}). More importantly, while the three molecules were associated with the outcome in the screening model, they showed no association with cancer severity. Additionally, creatinine has been reported to have a negative association with the presence of pancreatic cancer \citep{Boursi2017-cn, Dong2018-hj}. Our results {\colr obtained using} MtCLM are consistent with these findings, although it should be noted that these results were {\colr based on} serum creatinine. The association of REG1B with cancer severity disappeared when using the fused {\colr lasso} and group lasso regularizations. These results suggest that the two tasks, screening and severity prediction, had significantly different structures in Debernardi's dataset, making MtCLM a more suitable choice.

In interpreting these results, it is important to note that our defined category $Y=0$ includes not only healthy individuals but also cases of non-cancer diseases. As shown in \ref{app:realdata}, there are differences in the distribution of markers between healthy individuals and those with non-cancer diseases. 

\subsection{METABRIC Cohort Dataset}\label{app:PDAC}
Breast cancer is a common cancer among women, and its causes and treatments are widely investigated. The Molecular Taxonomy of Breast Cancer International Consortium (METABRIC) provides a database of genetic mutations and transcriptome profiles from over 2,000 breast cancer specimens collected from tumor banks in the UK and Canada. Using the METABRIC dataset, efforts {\colr have been} made to explore distinct subgroups related to clinical characteristics and prognosis of breast cancer \citep[e.g.,][]{Curtis2012-bs, Mukherjee2018-ee, Rueda2019-kl}.

At the time the METABRIC cohort was conducted, the progression of breast cancer was roughly classified into five stages, from 0 to 4, based on tumor size and lymph node metastasis. Among these, Stages 0 and 1 indicate no lymph node metastasis, while Stages 2 and above include metastasis to lymph nodes and other organs. Lymph node metastasis is used as a key indicator of cancer progression in many types of cancer, as its presence is known to increase the risk of recurrence and metastasis \citep{ACS-qz}.

In this section, as a demonstration of methodology, data analysis was conducted to {\colr predict the presence or absence of lymph node metastasis and stage}, based on genetic mutations and transcriptome profiles. The original dataset is publicly available on cBioPortal \citep{Cerami2012-zv, Gao2013-tb, De_Bruijn2023-bc}, but for this analysis, a preprocessed version available on Kaggle datasets \citep{Alharbi2020-rt} was used. The dataset includes 1,403 cases with non-missing stage information, including 479 cases without lymph node metastasis, 800 cases with lymph node metastasis (Stage 2), 115 cases with lymph node metastasis (Stage 3), and 9 cases with lymph node metastasis (Stage 4). For constructing the prediction model, 701 cases were randomly assigned to the training set, and 702 cases to the validation set.
The predictors included 489 types of mRNA levels and 92 gene mutations {\colr (out of 173) that occurred in at least 10 cases in the training data.} The tuning parameters were selected from $(0.001, 0.005, 0.01, 0.05, 0.1)$ based on 5-fold CV.

Table \ref{tab:METABRIC_predict} shows the results of evaluating predictions using each model on the validation set. The L1-penalized logistic regression, a benchmark for screening, showed a low F1 Score based on a probability threshold of 0.5 {\colr probably due to the high prevalence of lymph node metastasis}. CLM and NPCLM underperformed logistic regression in terms of AUC for screening and all measures for overall evaluation including severity prediction. As discussed later, the weak association between the predictors and severity {\colr differences in severity among cases with lymph node metastasis in this data led to overfitting}. Although NPCLM was flexible, its performance did not significantly improve. MtCLM outperformed both CLM and NPCLM and showed competitive performance {\colr compared to logistic regression for screening.} However, for severity prediction, the predicted values were almost independent of the actual values. In fact, for the severity prediction model, no variables were selected except for MtCLM (L1 + Fused), which non-specifically selected many predictors.

\begin{table}[]
    \centering
    \caption{Predictive performances of the comparative methods in the METABRIC dataset. The left two columns show the classification performance for the w/ or w/o lymph node metastasis, while the right three columns show the classification performance for all categories, including severity prediction.}
    \begin{tabular}{lccccc}\hline
 & \multicolumn{2}{c}{Screening} & \multicolumn{3}{c}{Overall}  \\
method & AUC & F1 & Accuracy & MAE & Kendall's tau \\\hline
Logistic Reg. (L1) & 0.627 & 0.054 & - & - & - \\
CLM (L1) & 0.522 & 0.339 & 0.406 & 0.691 & 0.023 \\
NPCLM (L1) & 0.548 & 0.378 & 0.479 & 0.571 & 0.063 \\
MtCLM (L1) & 0.619 & 0.047 & 0.551 & 0.457 & 0.043 \\
MtCLM (L1 + Fused) & 0.635 & 0.083 & 0.556 & 0.453 & 0.080 \\
MtCLM (L1 + Group) & 0.619 & 0.039 & 0.551 & 0.457 & 0.038 \\ \hline
    \end{tabular}
    \label{tab:METABRIC_predict}
\end{table}

\begin{figure}[htbp]
    \centering
    \includegraphics[width = 0.7\textwidth]{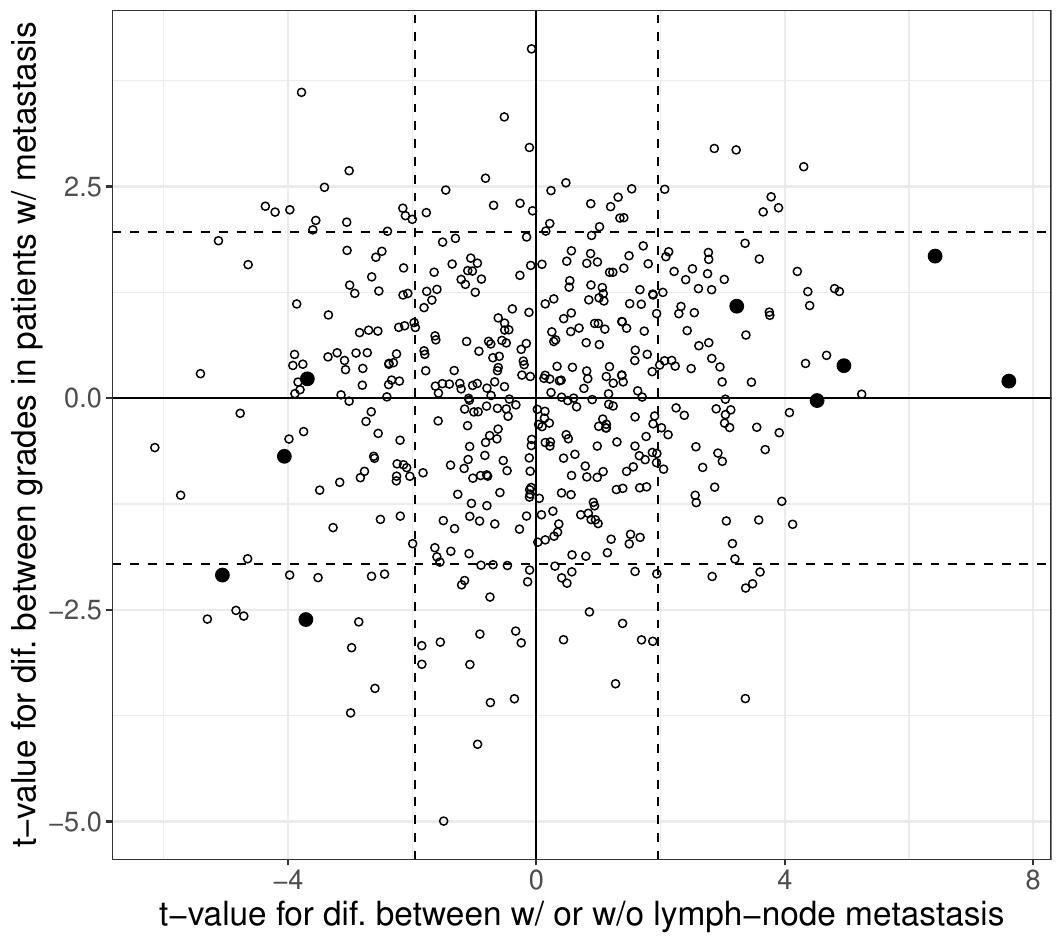}
    \caption{The relationship between the response and each gene expression level, and the {\colr variable selected by MtCLM (L1 + Group). The horizontal axis shows the difference in each gene between the presence and absence of lymph node metastasis,} and the vertical axis shows differences in mRNA levels between Stage 2 and higher stages for the cases with lymph node metastasis. The filled dots represent predictors selected by MtCLM (L1 + Group).}
    \label{fig:METABRIC_select}
\end{figure}

Figure \ref{fig:METABRIC_select} {\colr shows the association of 498 genes with the response variable in univariate analysis, and the variables selected in the MtCLM (L1 + Group) screening model.} Overall, variables strongly associated with the response variable in univariate analysis tended to be selected. For variables selected by other methods, refer to \ref{app:METABRIC}. 
Table \ref{tab:appMETABRIC} in \ref{app:METABRIC} presents the genes and mutations selected by each method. The twelve genes --- {\it BARD1, STAT5B, RBPJ, AURKA, CASP10, DIRAS3, GSK3B, RPS6KA2, SMAD2, RUNX1, HSD3B1, and FANCD2} (mutation) --- were selected even by the methods that yielded sparse results (the L1-penalized logistic reg., MtCLM with L1 penalty, and MtCLM with L1 and Group-lasso type penalty). Despite varying levels of supporting evidence, all selected genes have been previously linked to breast cancer, lending credibility to the biological relevance of the selection.
In particular, it is noteworthy that BARD1, a gene whose association with breast cancer has long been discussed \citep{Wu1996-pa}, was not selected by the CLM or the MtCLM with fused-lasso type penalty with cutoff, while it was selected by other methods. This observation suggests that, at least for the METABRIC data, a more flexible modeling approach may be more appropriate than CLM or fused-lasso type penalties that strongly impose similarity among regression coefficients.

\section{Discussion}

In this study, we addressed the combined problem of disease screening and severity prediction, recognizing that certain biomarkers may be associated not only with disease presence but also with disease severity. To solve this problem, we proposed a multi-task learning framework, MtCLM, which incorporates structured sparse regularization to leverage the shared structure between the two tasks. MtCLM is an ordinal regression model that offers greater flexibility than the conventional CLM. Unlike more flexible nonparametric alternatives such as NPCLM, it also maintains global validity and interpretability. We confirmed the effectiveness of MtCLM in terms of prediction accuracy and variable selection through simulation studies and demonstrated its practical utility with real biomedical data.

Our findings underscore the potential of multi-task learning in biomedical research, especially when tasks such as screening and severity prediction are expected to share underlying structures. Although multi-task learning is not yet widely applied in this domain, its value is evident in many clinical contexts. For example, hyperlipidemia is a known risk factor for both cardiovascular and cerebrovascular diseases \citep{Kopin2017-nu}, and HER2 gene amplification occurs in ovarian and gastric cancers as well as breast cancer \citep{Gravalos2008-mu}. These examples illustrate the opportunity to improve predictive modeling and biomarker discovery by sharing statistical strength across related tasks, particularly when data are limited. {\colrr If large-scale data are available, the MtCLM framework could be extended to address more complex and broadly applicable problems, as noted in Section \ref{sec31}. One possible direction is to incorporate a partially ordered PCM \citep{Zhang2012-so} into the model structure, adding a disease-specific branch to reflect the hierarchical relationship between disease presence and its severity. Another is to relax the linear predictor assumption and adopt more flexible functional forms, which would enable the framework to exploit commonalities not only between screening and severity prediction but also across different diseases. Such extensions would broaden the applicability of MtCLM and represent a promising direction for future research.}

Nevertheless, several limitations of the proposed method should be taken into account. First, as is common in structured regularization frameworks, MtCLM involves multiple tuning parameters, which can complicate parameter tuning. When tuning is performed via computationally intensive procedures such as cross-validation, runtime may become substantial. Future directions include the development of CV-free criteria such as information criteria, or the development of optimization algorithms tailored to MtCLM that are more efficient than general-purpose solvers like ADMM. Second, we observed relatively high false discovery rates in some scenarios, particularly when applying fused lasso penalties. While such penalties achieved high sensitivity, they tended to yield less sparse solutions. Although post hoc variable filtering based on coefficient magnitudes is a viable remedy, extensions to more aggressive sparsity-inducing penalties---such as $\ell_q$ regularization ($0 < q < 1$) \citep{Frank1993-au} or nonconvex penalties like SCAD \citep{Fan2001-qe}---may further enhance the model’s selection performance. However, these approaches would compromise the convexity of the loss function, a key advantage of MtCLM.

Despite these limitations, our results demonstrate that incorporating multi-task structures and domain-informed regularization can lead to more interpretable and effective predictive models in biomedical applications. We believe that MtCLM provides a flexible and extensible framework that opens new avenues for modeling complex clinical tasks in a statistically principled manner.


\section*{Acknowledgements}

Kazuharu Harada is partially supported by JSPS KAKENHI Grant Number 22K21286 and 25K21165.
Shuichi Kawano is partially supported by JSPS KAKENHI Grant Numbers JP23K11008, JP23H03352, JP23H00809, and JP25H01107.
Masataka Taguri is partially supported by JSPS KAKENHI Grant Number 24K14862.

\section*{Declaration of Generative AI and AI-assisted technologies in the writing process}
During the preparation of this work, the authors used ChatGPT 4/4o (OpenAI Inc.) in order to improve English writing. After using this tool/service, the authors reviewed and edited the content as needed and take full responsibility for the content of the publication.

\par


\newpage
\bibliographystyle{abbrvnat}
\bibliography{main}

\clearpage
\appendix

\setcounter{figure}{0}
\setcounter{table}{0}
\setcounter{equation}{0}
\renewcommand{\thesection}{Appendix \Alph{section}}
\renewcommand{\thefigure}{A\arabic{figure}}
\renewcommand{\thetable}{A\arabic{table}}
\renewcommand{\theequation}{A\arabic{equation}}
\renewcommand{\thetheorem}{A\arabic{theorem}}
\renewcommand{\thelemma}{A\arabic{lemma}}

\section{Further Illustrations for the Learning with Structural Sparse Regularization}\label{app:Sparsity}
\begin{figure}[ht]
    \centering
    \includegraphics[width = 0.9\textwidth]{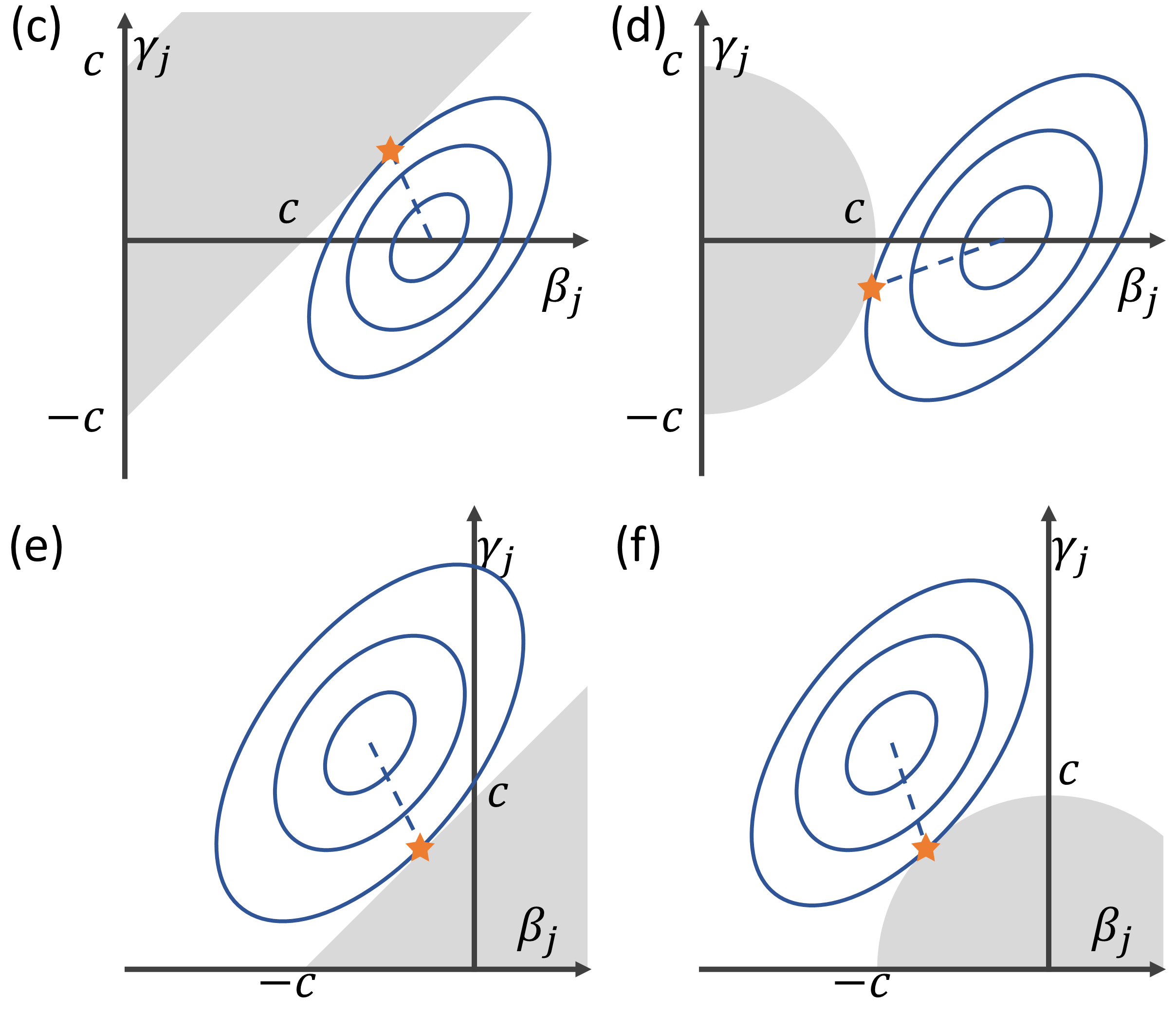}
    \caption{(Continues from Figure \ref{fig:sparse1}) Illustrations for the structural sparse penalties and the optimal solution regarding the regression coefficients of $X_j$. (Left) The inequality constraint of fused lasso type penalty (gray shading; $|\beta_j - \gamma_j|\le c$) and the log-likelihood function (blue contour lines). (Right) The inequality constraint of group lasso type penalty (gray shading; $\sqrt{\beta_j^2 + \gamma^2_j}\le c$) and the log-likelihood function. (Upper) the cases with $\beta_j > 0$ and $\gamma_j = 0$ being true. (Lower) the cases with $\beta_j > 0$ and $\gamma_j < 0$ being true. The star marks indicate the optimal solutions.}
    \label{fig:sparse_all}
\end{figure}

\clearpage
\section{Details of ADMM Algorithm}\label{app:ADMM}
\subsection{ADMM for Group lasso-type Estimation}\label{app:ADMM1}

In this subsection, we derive the ADMM algorithm for the problem \eqref{eq:optimG}. The optimization problem \eqref{eq:optimG} is equivalent to the following problem, which introduces a redundant parameter ${\bf B}\in\mathbb{R}^{p\times 2}$.
\begin{gather}\label{eq:optimADMM_G}
    \begin{array}{c}\displaystyle
        \min_{\alpha,\bs\zeta,\bs\Theta, {\bs B}}~-\frac{1}{n}\ell(\mathbf{Y},\mathbf{X}, \alpha,\bs\zeta,\bs\Theta) 
        + \lambda_{G}\sum_{j=1}^p\|{\bs b}_{j\cdot}\|_2 
        + \lambda_{11}\|{\bs b}_{\cdot 1}\|_1 + \lambda_{12}\|{\bs b}_{\cdot 2}\|_1\\
        \text{subject to} ~~~ \bs\Theta = {\bf B},
    \end{array}
\end{gather}
where ${\bs b}_{j\cdot}~(j=1,\ldots,p)$ are row vectors of ${\bf B}$.
The augmented Lagrangian of this problem is defined as

\begin{align}
    L(\alpha,\bs\zeta,\bs\Theta, {\bf B})
        =&~ -\frac{1}{n}\ell(\mathbf{Y},\mathbf{X}, \alpha,\bs\zeta,\bs\Theta) + \lambda_{G}\sum_{j=1}^p\|{\bs b}_{j\cdot}\|_2 + \lambda_{11}\|{\bs b}_{\cdot 1}\|_1 + \lambda_{12}\|{\bs b}_{\cdot 2}\|_1 \nonumber\\
        &~~~~ + \mathrm{tr}\{{\bf V}^T(\bs\Theta - {\bf B})\}
        + \frac{\mu}{2}\|\bs\Theta - {\bf B}\|^2_F, \label{eq:auglagG}
\end{align}
where $\bf V$ is a Lagrange multilier and $\mu$ is a tuning parameter for optimization.
In ADMM, the parameters are updated in sequence to minimize \eqref{eq:auglagG}. The Lagrange multipliers are updated by the gradient descent. Given the parameters of the previous step, the updating formulae are given below:
\begin{align}
    (\alpha^{t+1}, \bs\zeta^{t+1},\bs\Theta^{t+1})
        =&~ \argmin_{\bs\alpha,\bs\zeta,\bs\Theta}~
            L(\alpha,\bs\zeta,\bs\Theta, {\bf B}^{t}), \label{eq:upd1G}\\
    {\bf B}^{t+1}
        =&~ \argmin_{\bf B}~ \lambda_{G}\sum_{j=1}^p\|{\bs b}_{j\cdot}\|_2
            + \lambda_{11}\|{\bs b}_{\cdot 1}\|_1 + \lambda_{12}\|{\bs b}_{\cdot 2}\|_1 \nonumber\\
            &~~~~ + \mathrm{tr}\{{\bf V}^{tT}(\bs\Theta^{t+1} - {\bf B})\}
            + \frac{\mu}{2}\|\bs\Theta^{t+1} - {\bf B}\|_F^2, \label{eq:upd2G}\\
    {\bf V}^{t+1} 
        =&~ {\bf V}^{t} + \mu(\bs{\Theta}^{t+1} - {\bf B}^{t+1}), \label{eq:upd3G}
\end{align}
The small problem \eqref{eq:upd1G} does not have an explicit solution, so it has to be solved by an iterative algorithm. Since the target function of \eqref{eq:upd1G} is convex, it can be solved using an off-the-shelf solver. The problems \eqref{eq:upd2G} can be solved in the following 2-step procedure \cite{Simon2013-sc, Tugnait2021-lh}:
\begin{align*}
    {\bs b}_{\cdot 1}^{\dag} 
        =&~ S({\bs\beta}^{t+1} + \mu^{-1}{\bs v}_{\cdot 1}^{t}, \mu^{-1}\lambda_{11}),\\
    {\bs b}_{\cdot 2}^{\dag} 
        =&~ S({\bs\gamma}^{t+1} + \mu^{-1}{\bs v}_{\cdot 2}^{t}, \mu^{-1}\lambda_{12}),
\end{align*}
and
\begin{align*}
    {\bs b}_{j\cdot}^{t+1} 
        =&~ S_G({\bs b}_{j\cdot}^{\dag}, \sqrt{2}\mu^{-1}\lambda_G)~~~(j=1,\ldots,p),
\end{align*}
where $S_G(\bs z,\xi) = (1-\frac{\xi}{\|\bs z\|_2})_+{\bs z}$ is the soft-thresholding operator for a group lasso-type penalty, and ${\bs b}_{j\cdot}^{\dag}$ is the $j$th row vector of ${\mathbf{B}}^{\dag}$.
Repeat steps \eqref{eq:upd1G} through \eqref{eq:upd3G} until an appropriate convergence criterion is met to obtain the final estimate. 

\subsection{Gradient of the Augmented Langrangian}\label{app:ADMM2}
We derive the gradient of the augmented Lagrangian for the fused lasso-type problem.
Note that the derivative of $\sigma(u)$ is $\sigma^{[1]}(u) = \sigma(u)\{1 - \sigma(u)\}$, which is also the density function of the logistic distribution. Each gradient is given as follows:
\begin{align*}
    \frac{\partial L}{\partial \alpha} 
        = -\frac{1}{n}\frac{\partial \ell}{\partial \alpha} 
        = -\frac{1}{n}\frac{\partial \log \mathcal{L}_1}{\partial \alpha}
        = -\frac{1}{n}\sum_{i=1}^n \left[
            {\bf 1}(Y_i = 0) - \sigma\{\alpha + X_i\bs\beta\}
        \right],
\end{align*}

\begin{align*}
    \frac{\partial L}{\partial \bs\beta} 
        =&~ -\frac{1}{n}\frac{\partial}{\partial \bs\beta}\left\{
            \log \mathcal{L}_1 + {\bs u}^T(\bs\Theta{\bs d} - \bs a) 
            + \frac{\mu_F}{2}\|\bs\Theta{\bs d} - \bs a\|^2_2 
            + \mathrm{tr}\left\{
                \mathbf{V}^T(\bs\Theta - \mathbf{B})
            \right\}
            + \frac{\mu_1}{2}\|\bs\Theta - \mathbf{B}\|^2_F
        \right\}\\
        =&~ -\frac{1}{n}\sum_{i=1}^n \left[
            {\bf 1}(Y_i = 0) - \sigma\{\alpha + X_i\bs\beta\}
        \right]X_i + {\bs u} + \mu_F(\bs\Theta{\bs d} - {\bs a}) + {\bs v}_1 + \mu_1({\bs\beta} - {\bs b}_{\cdot 1}),
\end{align*}

\begin{align*}
    \frac{\partial L}{\partial \zeta_k} 
        =&~ -\frac{1}{n}\frac{\partial \log \mathcal{L}_2}{\partial \zeta_k} \\
        =&~ -\frac{1}{n}\sum_{i=1}^n \left[
            \frac{{\bf 1}(Y_i = k)\sigma^{[1]}(\zeta_k + X_i^T\bs\gamma)}{\sigma(\zeta_k + X_i^T\bs\gamma) - \sigma(\zeta_{k-1} + X_i^T\bs\gamma)} -\frac{{\bf 1}(Y_i = k + 1)\sigma^{[1]}(\zeta_k + X_i^T\bs\gamma)}{\sigma(\zeta_{k+1} + X_i^T\bs\gamma) - \sigma(\zeta_{k} + X_i^T\bs\gamma)} 
        \right],
\end{align*}

\begin{align*}
    \frac{\partial L}{\partial \bs\gamma} 
        =&~ -\frac{1}{n}\frac{\partial}{\partial \bs\gamma} \left\{
            \log \mathcal{L}_2 +  {\bs u}^T(\bs\Theta{\bs d} - \bs a)
            + \frac{\mu}{2}\|\bs\Theta{\bs d} - \bs a\|^2_2
            + \mathrm{tr}\left\{
                \mathbf{V}^T(\bs\Theta - \mathbf{B})
            \right\}
            + \frac{\mu_1}{2}\|\bs\Theta - \mathbf{B}\|^2_F
        \right\}\\
        =&~ -\frac{1}{n}\sum_{i=1}^n\sum_{k=1}^K \left[
            {\bf 1}(Y_i = k)X_i\frac{\sigma^{[1]}(\zeta_k + X_i^T\bs\gamma) - \sigma^{[1]}(\zeta_{k-1} + X_i^T\bs\gamma)}{\sigma(\zeta_k + X_i^T\bs\gamma) - \sigma(\zeta_{k-1} + X_i^T\bs\gamma)}
        \right] \\
        &~~~~~ - {\bs u} - \mu(\bs\Theta{\bs d} - {\bs a}) + {\bs v}_2 + \mu_1({\bs\gamma} - {\bs b}_{\cdot 2}).
\end{align*}
Since $\zeta_0 = -\infty$ and $\zeta_K = \infty$, we have $\sigma^{[1]}(\zeta_0 + X_i^T\bs\gamma) = \sigma^{[1]}(\zeta_K + X_i^T\bs\gamma) = 0$.

For the group lasso-type augmented Lagrangian, some modifications are needed. Let $\bs v_1$ and $\bs v_2$ be the column vectors of $\mathbf{V}$, and then the gradients are
\begin{align*}
    \frac{\partial L}{\partial \bs\beta} 
        =&~ -\frac{1}{n}\frac{\partial }{\partial \bs\beta}\left\{
            \log \mathcal{L}_1 + \mathrm{tr}\left\{
                \mathbf{V}^T(\bs\Theta - \mathbf{B}) 
            \right\}
            + \frac{\mu}{2}\|\bs\Theta - \mathbf{B}\|^2_F
        \right\}\\
        =&~ -\frac{1}{n}\sum_{i=1}^n \left[
            {\bf 1}(Y_i = 0) - \sigma\{\alpha + X_i\bs\beta\}
        \right]X_i + {\bs v}_1 + \mu(\bs\beta - {\bs b}_{\cdot 1}), \\
    \frac{\partial L}{\partial \bs\gamma} 
        =&~ -\frac{1}{n}\frac{\partial}{\partial \bs\gamma} \left\{
            \log \mathcal{L}_2 + \mathrm{tr}\left\{
                \mathbf{V}^T(\bs\Theta - \mathbf{B}) 
            \right\}
            + \frac{\mu}{2}\|\bs\Theta - \mathbf{B}\|^2_F
        \right\}\\
        =&~ -\frac{1}{n}\sum_{i=1}^n\sum_{k=1}^K \left[
            {\bf 1}(Y_i = k)X_i\frac{\sigma^{[1]}(\zeta_k + X_i^T\bs\gamma) - \sigma^{[1]}(\zeta_{k-1} + X_i^T\bs\gamma)}{\sigma(\zeta_k + X_i^T\bs\gamma) - \sigma(\zeta_{k-1} + X_i^T\bs\gamma)}
        \right] + {\bs v}_2 + \mu(\bs\gamma - {\bs b}_{\cdot 2}).
\end{align*}

\clearpage
\subsection{Pseudocode}\label{app:ADMM3}
In this section, we present pseudocode for the ADMM algorithm used to estimate MtCLM. While several stopping criteria are possible, our implementation adheres to the guidelines in Section 3.3.1 of \citep{Boyd2011-ak}.

\begin{algorithm}
\caption{ADMM for MtCLM with L1 + Fused-lasso-type Penalty}
\begin{algorithmic}[1]
    \State \textbf{Input:} Data: $\{X_i,Y_i\}_{i=1}^n$, Tuning parameters for regularization: $\lambda_{11}, \lambda_{12}, \lambda_F$, Optimization parameters: $\mu_F, \mu_1$.
    \State \textbf{Initialize:} $\alpha^{0}, \bs\zeta^{0}, \Theta^{0}, {\bs a}^{0}, \mathbf{B}^{0}$.
    \While{not converged}
        \State $(\alpha^{t+1}, \bs\zeta^{t+1},\bs\Theta^{t+1}) =  \argmin_{\bs\alpha,\bs\zeta,\bs\Theta}~  L(\alpha,\bs\zeta,\bs\Theta, {\bs a}^{t}, \mathbf{B}^{t})$ \text{via an off-the-shelf optimizer.}
        \State $\bs{a}^{t+1} = S(\bs{\Theta}^{t+1}{\bs d} + \mu_F^{-1}\bs{u}^{t}, \mu_F^{-1}\lambda_F)$
        \State $\mathbf{\bs b}_{\cdot 1}^{t+1} 
        = S(\bs{\beta}^{t+1} + \mu_1^{-1}{\bs v}_{\cdot 1}^{t+1}, \mu_1^{-1}\lambda_{11})$
        \State $\mathbf{\bs b}_{\cdot 2}^{t+1} 
        = S(\bs{\gamma}^{t+1} + \mu_1^{-1}{\bs v}_{\cdot 2}^{t+1}, \mu_1^{-1}\lambda_{12})$
        \State $\bs{u}^{t+1} 
        = \bs{u}^{t} + \rho(\bs{\Theta}^{t+1}{\bs d} - \bs{a}^{t+1})$
        \State $\mathbf{V}^{t+1} 
        = \mathbf{V}^{t} + \rho(\bs{\Theta}^{t+1} - \mathbf{B}^{t+1})$
        \If{converged}
            \Return $\Theta^{t+1}$ 
        \EndIf
    \EndWhile
\end{algorithmic}
\end{algorithm}

\begin{algorithm}
\caption{ADMM for MtCLM with L1 + Group-lasso-type Penalty}
\begin{algorithmic}[1]
    \State \textbf{Input:} Data: $\{X_i,Y_i\}_{i=1}^n$, Tuning parameters for regularization: $\lambda_{11}, \lambda_{12}, \lambda_G$, Optimization parameters: $\mu$.
    \State \textbf{Initialize:} $\alpha^{0}, \bs\zeta^{0}, \Theta^{0}, {\bs a}^{0}, \mathbf{B}^{0}$.
    \While{not converged}
        \State $(\alpha^{t+1}, \bs\zeta^{t+1},\bs\Theta^{t+1})
        = \argmin_{\bs\alpha,\bs\zeta,\bs\Theta}~L(\alpha,\bs\zeta,\bs\Theta, {\bf B}^{t})$ \text{via an off-the-shelf optimizer.}
        \State $\bs{a}^{t+1} = S(\bs{\Theta}^{t+1}{\bs d} + \mu_F^{-1}\bs{u}^{t}, \mu_F^{-1}\lambda_F)$
        \State ${\bf B}^{t+1}= \argmin_{\bf B}~ \lambda_{G}\sum_{j=1}^p\|{\bs b}_{j\cdot}\|_2 + \lambda_{11}\|{\bs b}_{\cdot 1}\|_1 + \lambda_{12}\|{\bs b}_{\cdot 2}\|_1  + \mathrm{tr}\{{\bf V}^{tT}(\bs\Theta^{t+1} - {\bf B})\} + \frac{\mu}{2}\|\bs\Theta^{t+1} - {\bf B}\|_F^2$
        
        \State ${\bf V}^{t+1} = {\bf V}^{t} + \mu(\bs{\Theta}^{t+1} - {\bf B}^{t+1})$
        \If{converged}
            \Return $\Theta^{t+1}$ 
        \EndIf
    \EndWhile
\end{algorithmic}
\end{algorithm}

\clearpage
\subsection{Convergence}\label{app:ADMM4}
According to Section 3.2 of \citet{Boyd2011-ak}, the ADMM applied to
\begin{gather*}
    \min_{x,z} f(x) + g(z)~~~\text{subject to}~~~Ax+Bz = c
\end{gather*}
is guaranteed to converge under Assumptions 1 and 2.
\begin{assumption}[\cite{Boyd2011-ak}]
The (extended-real-valued) functions $f:\mathbb{R}^n\rightarrow\mathbb{R}\cup\{+\infty\}$ and $g: \mathbb{R}^m\rightarrow\mathbb{R}\cup\{+\infty\}$ are closed, proper, and convex.
\end{assumption}

\begin{assumption}[\cite{Boyd2011-ak}]
The unaugmented Lagrangian $L_0$ has a saddle point.
\end{assumption}

In our formulations \eqref{eq:optimADMM_F} and \eqref{eq:optimADMM_G}, $f$ is the (negative) log-likelihood of a logistic-regression component combined with that of an ordinal CLM for patients, $g$ is the regularization term, and the equality constraint enforces equality between the original and redundant parameters.

First, we verify Assumption 1. Both negative log-likelihood terms are convex and continuous \citep{Burridge1981-mn,Pratt1981-oa,Agresti2015-ez}. The regularization terms are likewise convex and continuous. The sum of closed convex functions is closed and convex, hence $f+g$ is closed and convex. Next, we verify that the objective function is proper.
Because the linear constraint simply forces the redundant parameters to equal the original ones, a feasible point clearly exists.  For instance, choosing
$$
\alpha = 0,\quad \boldsymbol\beta = \boldsymbol\gamma = \mathbf 0,\quad
\{\zeta_k\}_{k=0}^K = \bigl(-\infty,\,1,\,2,\ldots,\,K-1,\,\infty\bigr)
$$
satisfies the constraint and yields a finite objective value.
The (negative) log-likelihood for logistic regression contains terms of the form $-\log\sigma(u)$; since $\sigma(u)\in(0,1)$, these terms are always strictly positive and never diverge to $-\infty$.  Likewise, each CLM log-likelihood term
$$
-\log\!\bigl\{\sigma(\zeta_k+\cdots)-\sigma(\zeta_{k-1}+\cdots)\bigr\}
$$
is bounded below by $0$ whenever $\zeta_k \ge \zeta_{k-1}$. The regularization term is non-negative and finite whenever the parameters are finite.
Hence, the objective attains a finite value at some point and never attains $-\infty$; the function is proper, and Assumption 1 is satisfied for our optimization problem.

Next, we verify Assumption 2. The overall objective is convex in all parameters, and the constraint set is non-empty, so the Slater condition holds. Then, the problem enjoys strong duality; equivalently, the unaugmented Lagrangian possesses a saddle point \citep{Boyd2016-cd}. Therefore, Assumption 2 holds.

Consequently, we have established that the ADMM algorithm for the proposed method is guaranteed to converge in terms of residual, objective function, and the dual variable.

Below, we present a numerical experiment under a simple setting to empirically verify the convergence of ADMM for the proposed method.
We considered three regularization schemes: L1 only, L1 + Fused, and L1 + Group. 
{\colg Under the ``Similar'' scenario in Section 5, we tracked the augmented Lagrangian at every ADMM iteration. The sequence quickly settled to a constant, indicating that the iterates approached a saddle point of the augmented Lagrangian and that the original problem was being solved appropriately.}
The regularization parameters were set to $\lambda_{11} = \lambda_{12} = 0.05$ for all methods. For the L1 + Fused method, we used $\lambda_F = 0.01$, $\lambda_G = 0$;
for the L1 + Group method, $\lambda_F = 0$, $\lambda_G = 0.01$. The step size for updating the Lagrange multipliers was set to 1.

\begin{figure}[htbp]
    \centering
    \includegraphics[width = \textwidth]{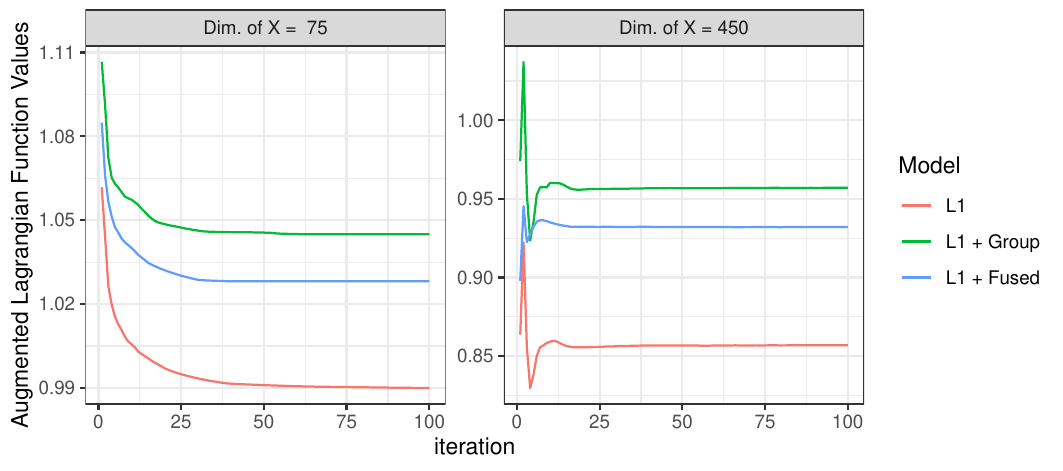}
    \caption{Empirical verification of the convergence of the augmented Lagrangian in ADMM for MtCLM}
    \label{fig:convergence}
\end{figure}

\clearpage
\section{Details on the Numerical Experiments}\label{app:experiments}
\subsection{Simulation Models}
The datasets for the numerical experiments are generated from the CLM and MtCLM models. Specifically, the ordinal response $Y$ is generated by discretizing continuous variables $Y^*$ and $Y^{**}$ into four categories (for the scenario of CLM, we only generate $Y^*$). $Y^*$ and $Y^{**}$ are generated by adding error terms that follow a standard logistic distribution to the linear combinations of the predictors. The discretization thresholds are set so that the four categories comprise 50\%, 16.7\%, 16.7\%, and 16.7\% of the data, respectively. All predictors are generated from a $p$-dimensional multivariate normal distribution, where they are mutually independent. The dimension is in ${75, 150, 300, 450}$. In the case of CLM, 10 out of the $p$ regression coefficients were set to non-zero, while in the case of MtCLM, 10 out of the total $p$ regression coefficients for each task are set to non-zero. The absolute values of the non-zero regression coefficients were drawn from a uniform distribution over $(0.75, 1.25)$. The selection and sign of the non-zero regression coefficients depend on the following scenarios. The true regression coefficients for the signal predictors are set to $\pm 0.5$ in low-dimensional settings, and set to $\pm 1.0$ in high-dimensional settings. The signs and zero/non-zero patterns are different scenario by scenario as follows.

\paragraph{Scenario 1: Parallel}
The simulation model is a parallel CLM. Of the 10 non-zero coefficients, the first 5 are positively associated with $ Y*$, and the remaining 5 are negatively associated with $ Y*$.

\paragraph{Scenario 2: Identical}
The simulation model is a MtCLM. All relevant predictors have regression coefficients with the same signs in screening and severity prediction.
Among $10$ non-zero coefficients for both tasks, the first 5 are positively associated with $Y^*$, and the remaining 5 are negatively associated with $Y^*$.

\paragraph{Scenario 3: Almost Inverse}
The simulation model is an MtCLM. The relevant predictors are common in both screening and severity prediction, but the signs of the regression coefficients are almost inverse for each task. Specifically, among the 10 non-zero coefficients for the screening model, the first 5 are positive, and the remaining 5 are negative. In contrast, the signs are inverse in the severity model, except for the first and sixth coefficients.

\paragraph{Scenario 4: Similar}
The simulation model is an MtCLM. The screening and severity models share many relevant predictors, but some are different. 
Specifically, among the 10 non-zero coefficients for the screening model, the first 5 coefficients are positive, and the remaining 5 are negative. For the severity model, the same predictors are associated with the response in the same direction, except for the 4th and 9th. Instead of the 4th and 9th predictors, the 11th and 12th predictors are associated with the response.

\paragraph{Scenario 5: Almost Independent}
The simulation model is an MtCLM. Only two predictors are shared between both tasks. Specifically, among the 10 non-zero coefficients for the screening model, the first 5 coefficients are positive, and the remaining 5 are negative. For the severity model, only the 1st and 6th predictors are associated with the response in the same direction, whereas the 11th to 14th predictors are positively associated with severity, and the 15th to 18th predictors are negatively associated with severity.

\subsection{Evaluation Measures}
We use the following evaluation measures to evaluate the predictive performance. The ROC-AUC and the F1 score are used to evaluate performance in screening, and the accuracy, MAR, and Kendall's tau are used to evaluate performance in the combined task of screening and severity prediction.
\begin{itemize}
    \item ROC-AUC: area under the receiver operating characteristic curve. We simply refer to it as AUC. This takes values in {\colg $[0.5, 1.0]$} and it measures the overall performance for the binary classification of a continuous indicator, taking into account sensitivity and specificity on various cutoff values.
    \item F1 Score: the harmonic mean of the positive predictive value (PPV, or Precision) and the sensitivity (or Recall), which measures the overall performance for the binary classification of a specific rule.
    \item Accuracy: the proportion of the accurate prediction across all levels of the ordinal response.
    \item MAE: the mean absolute error for the ordinal response.
    \item Kendall's tau: a statistic that measures the ordinal association between two quantities and is one measure of concordance. A value of $-1$ indicates complete disagreement, 0 indicates no association, and 1 indicates complete agreement.
\end{itemize}

To evaluate the variable selection performance, we use the following measures.
\begin{itemize}
    \item F1 Score
    \item False Discovery Rate (FDR): False Positive / Predicted Positive.
    \item Sensitivity: True Positive / Positive.
    \item Specificity: True Negative / Negative.
\end{itemize}

\clearpage
\section{Additional Results for Numerical Experiments}\label{app:results}
\subsection{Experimental Results for $p = 75, 150, 300$}
In addition to the setting with $N=300, p=450$ presented in the main text, comparisons were also conducted for $p \in \{75, 150, 300\}$.

\begin{figure}[htbp]
    \centering
    \includegraphics[width = \textwidth]{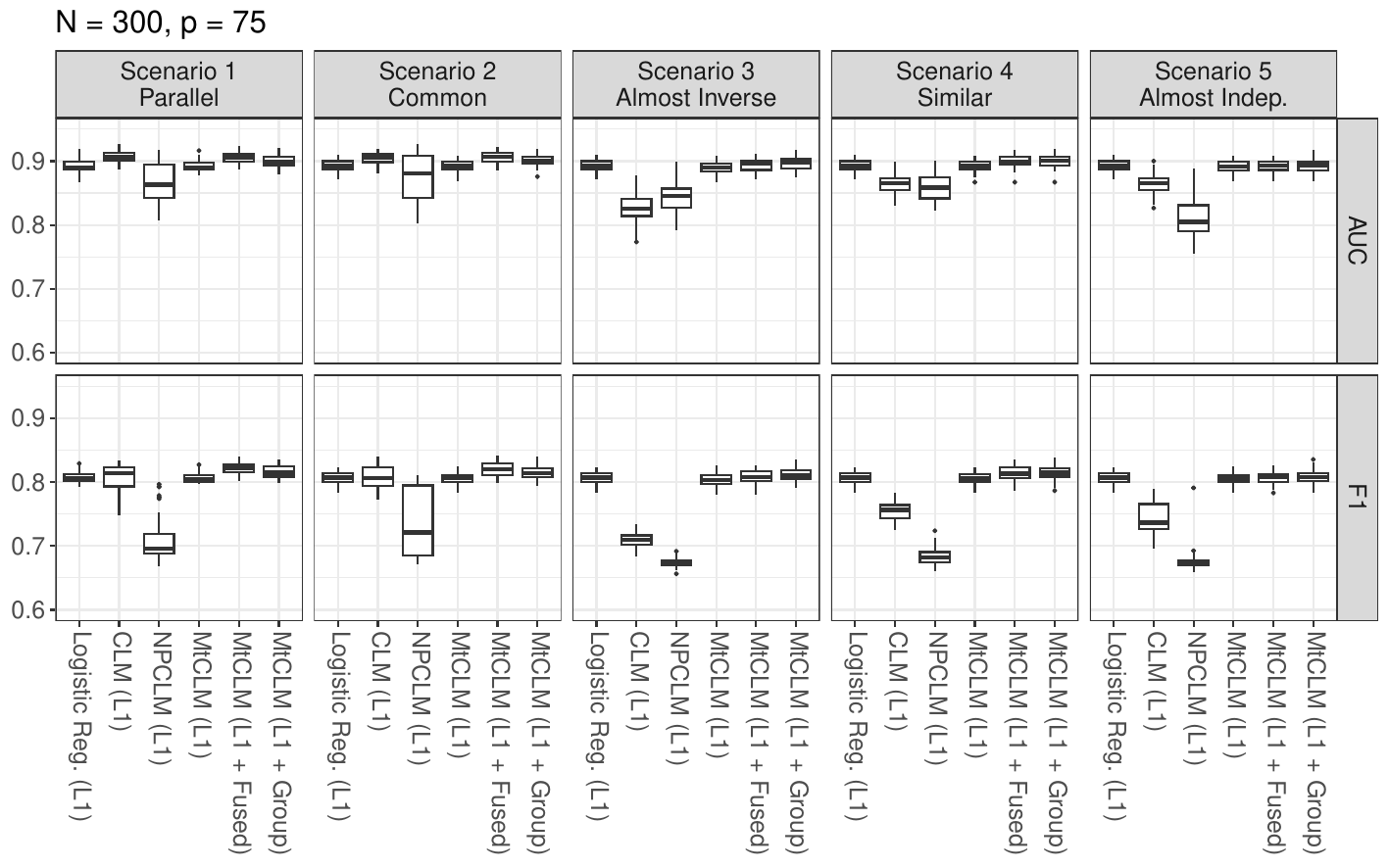}
    \caption{Comparison of the proposed and existing methods for the screening (0/1 classification) with 75-dimensional predictors.}
    \label{fig:performance_in_diag_075}
\end{figure}

\begin{figure}[htbp]
    \centering
    \includegraphics[width = \textwidth]{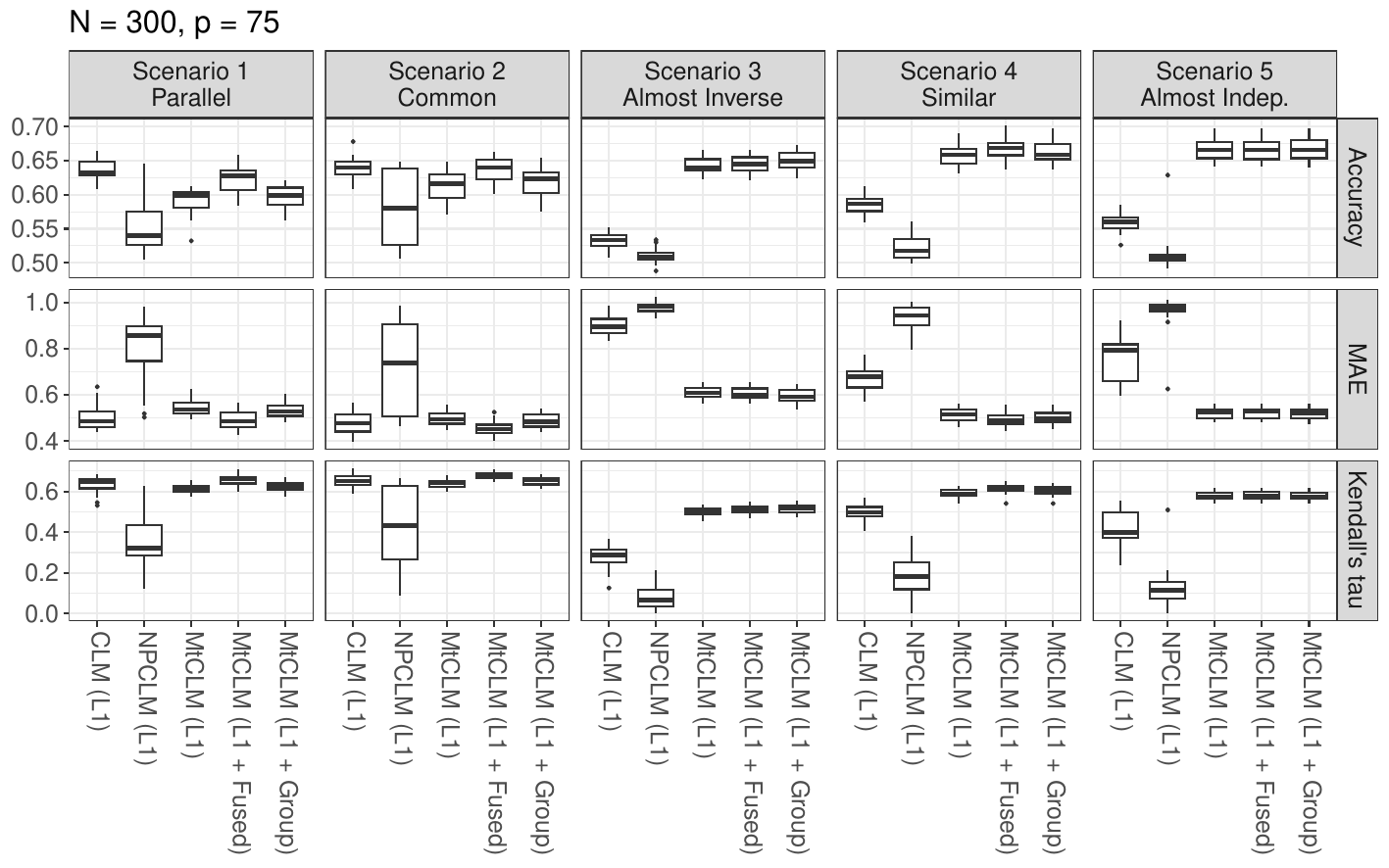}
    \caption{Comparison of the proposed and existing methods for the joint task of screening and severity prediction with 75-dimensional predictors.}
    \label{fig:performance_in_joint_075}
\end{figure}

\begin{figure}[htbp]
    \centering
    \includegraphics[width = \textwidth]{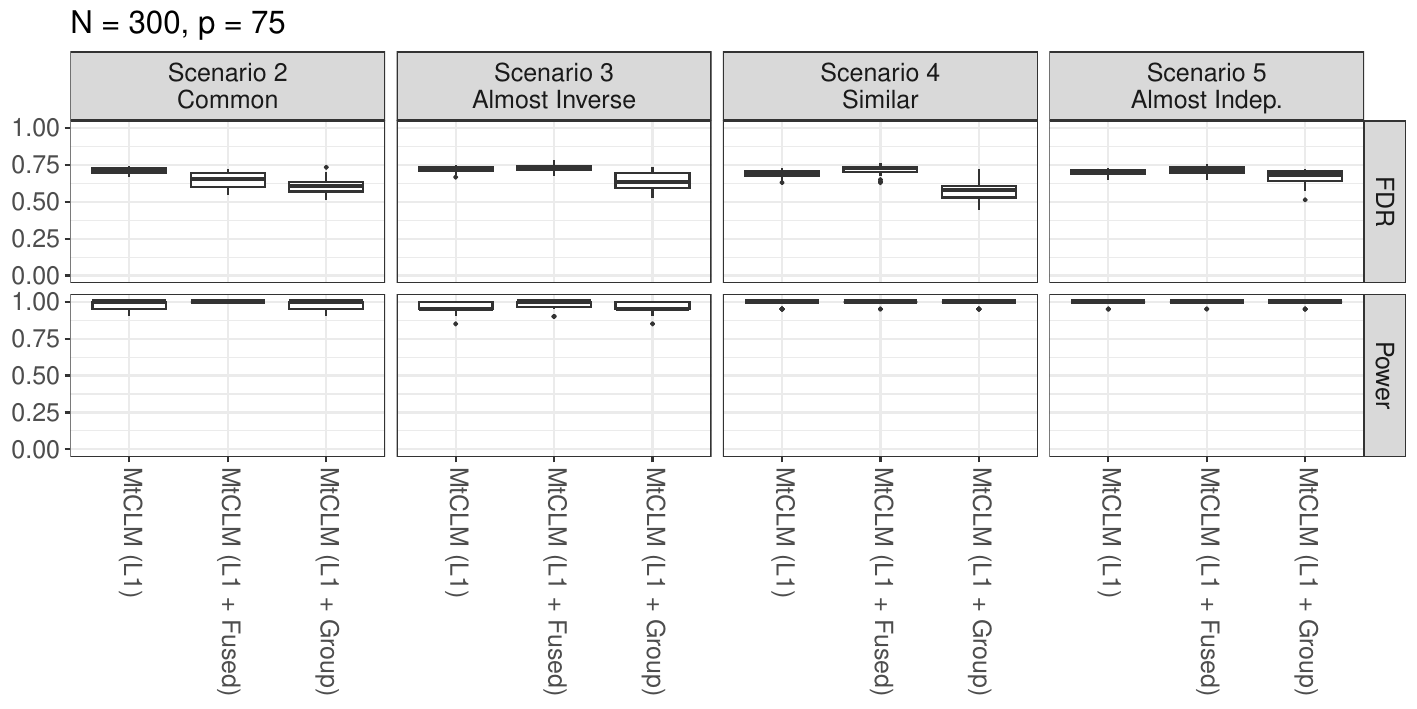}
    \caption{Comparison of the proposed methods in variable selection among 75-dimensional predictors.}
    \label{fig:performance_in_selection_075}
\end{figure}

\begin{figure}[htbp]
    \centering
    \includegraphics[width = \textwidth]{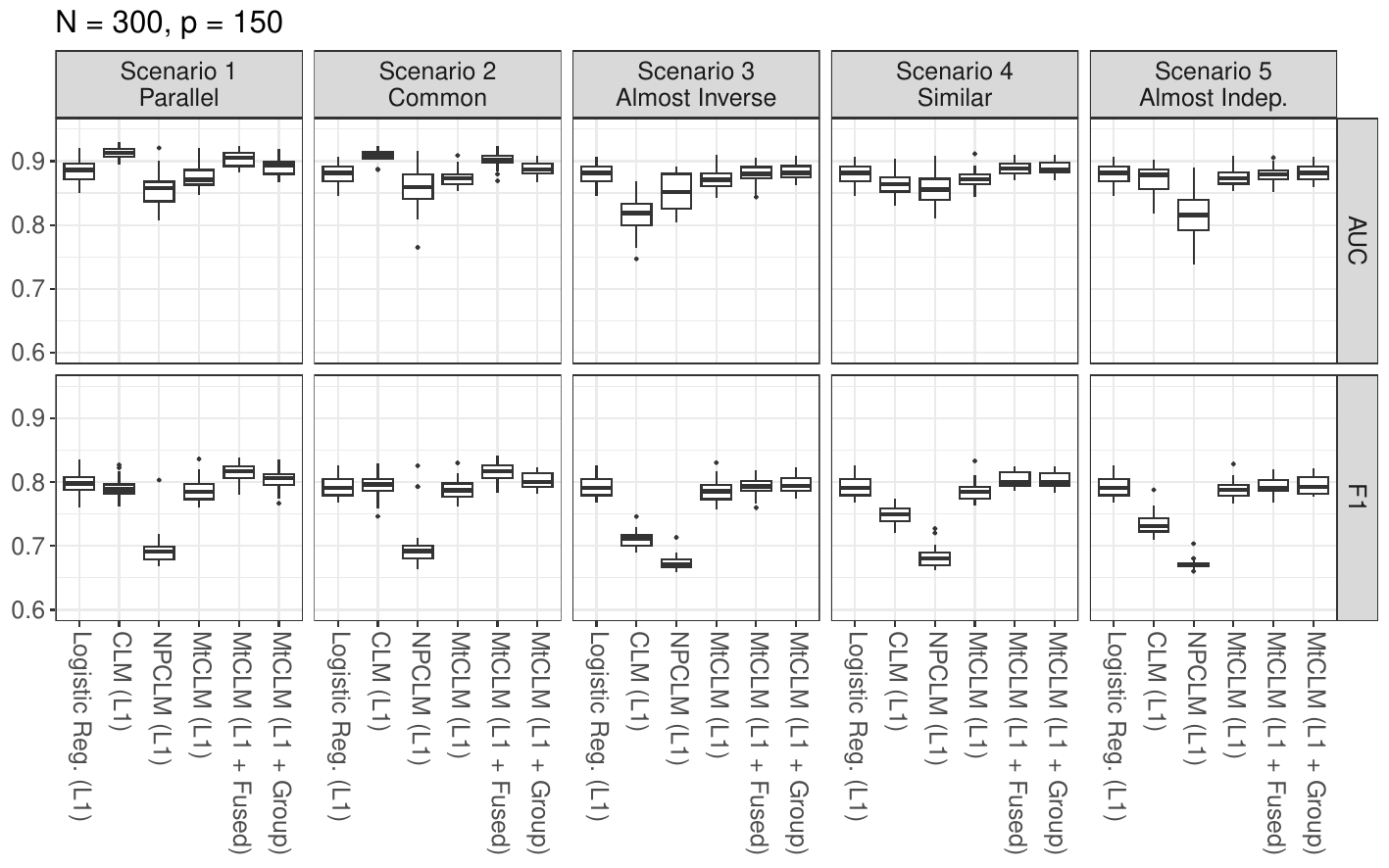}
    \caption{Comparison of the proposed and existing methods for the screening (0/1 classification) with 150-dimensional predictors.}
    \label{fig:performance_in_diag_150}
\end{figure}

\begin{figure}[htbp]
    \centering
    \includegraphics[width = \textwidth]{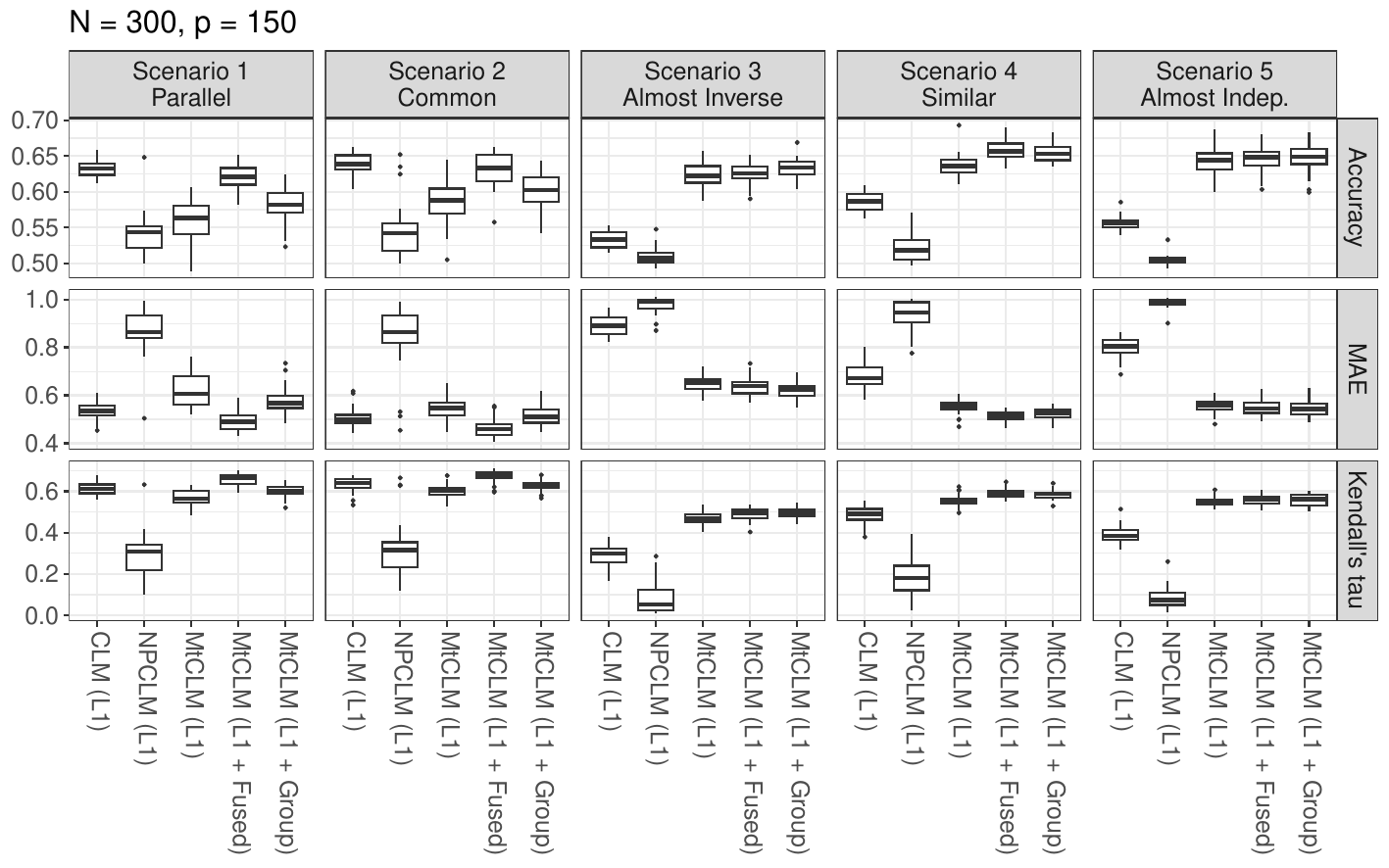}
    \caption{Comparison of the proposed and existing methods for the joint task of screening and severity prediction with 150-dimensional predictors.}
    \label{fig:performance_in_joint_150}
\end{figure}

\begin{figure}[htbp]
    \centering
    \includegraphics[width = \textwidth]{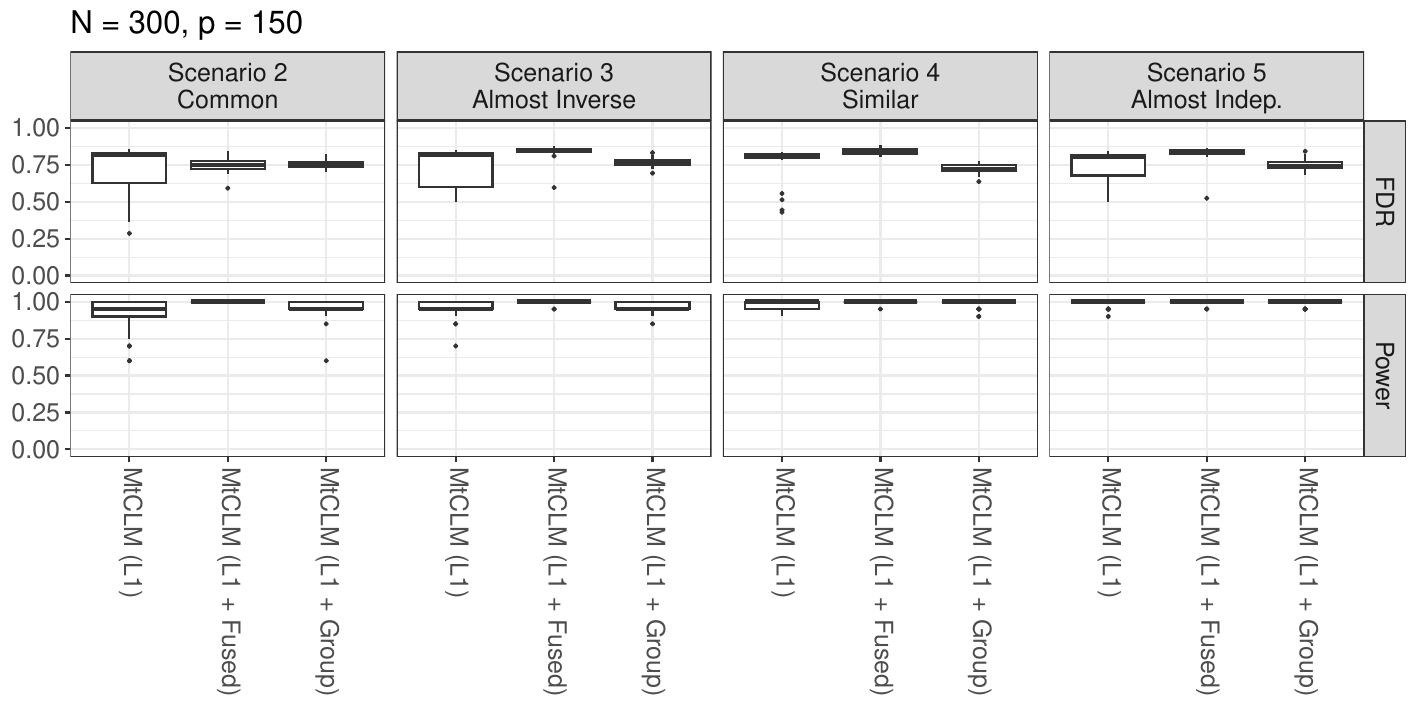}
    \caption{Comparison of the proposed methods in variable selection among 150-dimensional predictors.}
    \label{fig:performance_in_selection_150}
\end{figure}

\begin{figure}[htbp]
    \centering
    \includegraphics[width = \textwidth]{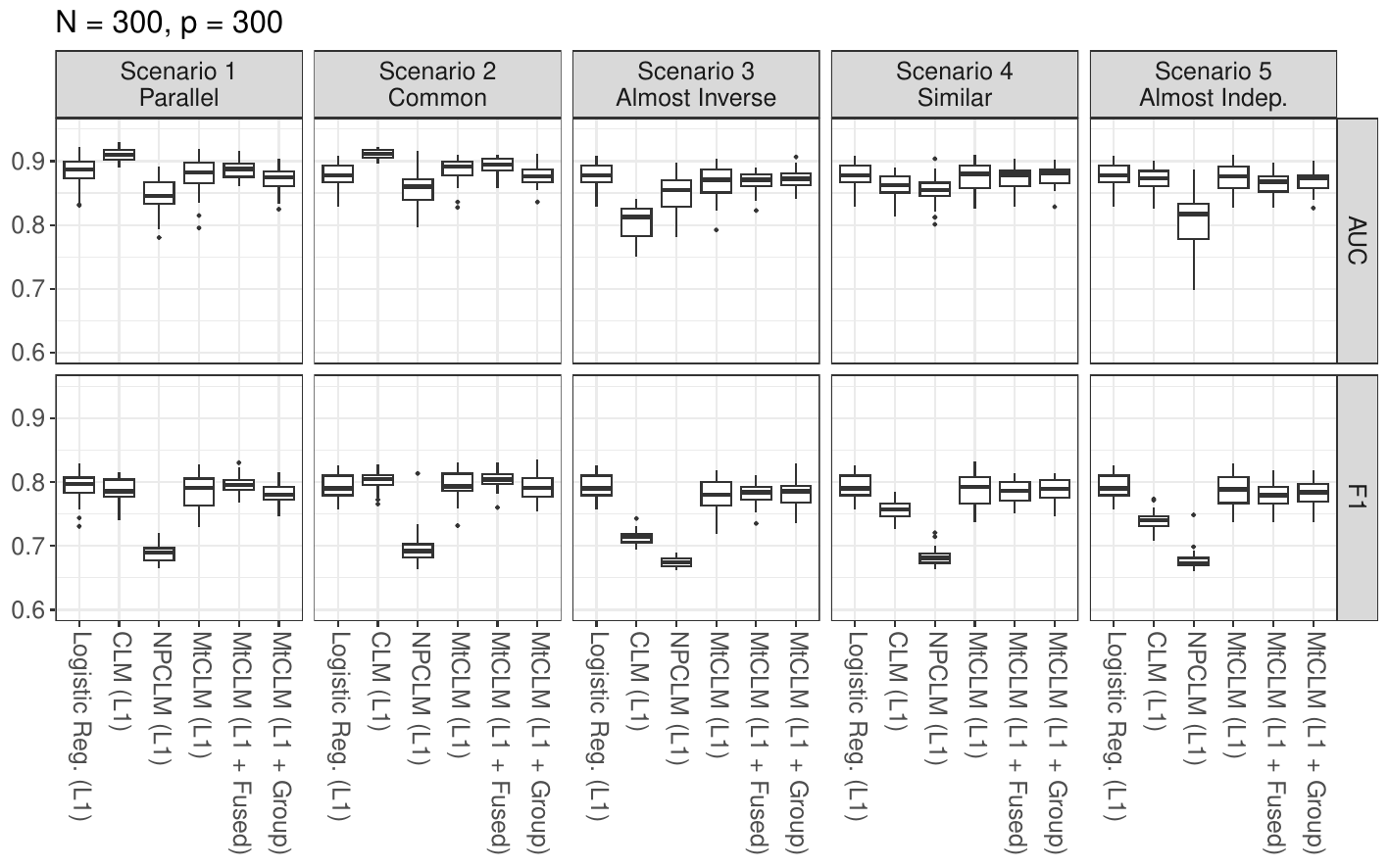}
    \caption{Comparison of the proposed and existing methods for the screening (0/1 classification) with 300-dimensional predictors.}
    \label{fig:performance_in_diag_300}
\end{figure}

\begin{figure}[htbp]
    \centering
    \includegraphics[width = \textwidth]{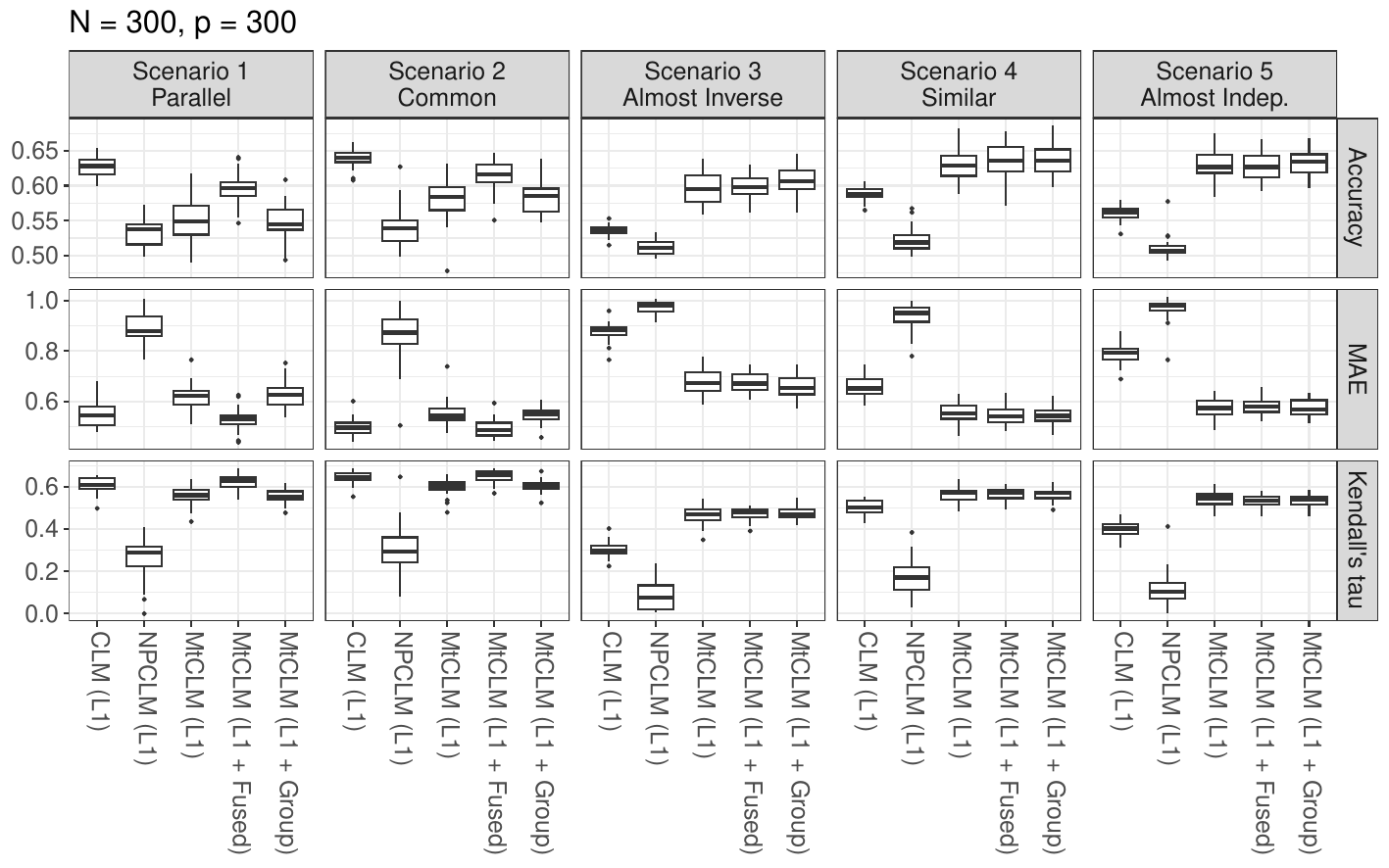}
    \caption{Comparison of the proposed and existing methods for the joint task of screening and severity prediction with 300-dimensional predictors.}
    \label{fig:performance_in_joint_300}
\end{figure}

\begin{figure}[htbp]
    \centering
    \includegraphics[width = \textwidth]{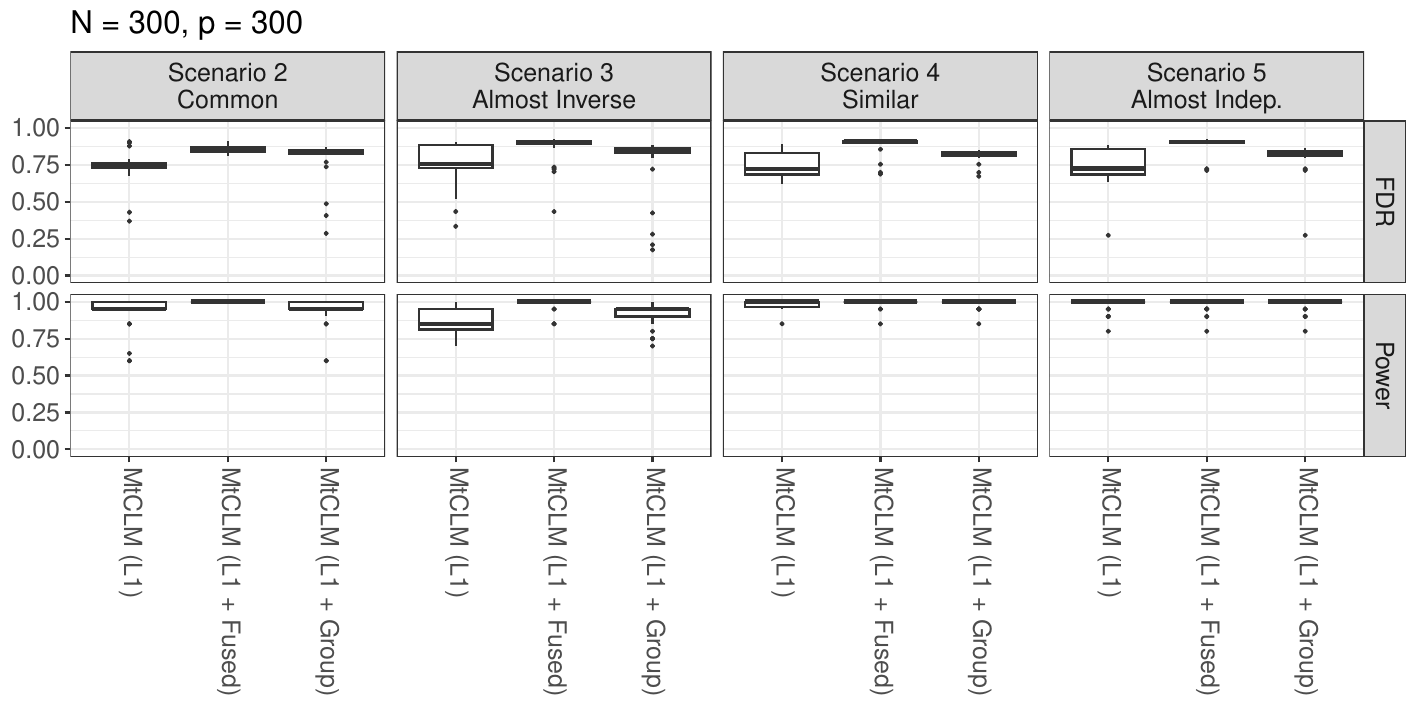}
    \caption{Comparison of the proposed methods in variable selection among 300-dimensional predictors.}
    \label{fig:performance_in_selection_300}
\end{figure}

\clearpage
\subsection{Experimental Results with Correlated Predictors}
For each of the five scenarios, we compared the performance of the proposed and existing methods under the setting where the predictors \(X\) follow a multivariate normal distribution with a Toeplitz correlation structure—that is, the correlation between the \(i\)th and \(j\)th predictors is given by \(\rho^{|i - j|}\), where \(\rho\in\{0,0.3,0.6,0.9\}\) is a non-negative constant. The dimension of \(X\) was set to two values: \(p \in \{75, 450\}\).

Throughout the experiments in this section, we observed that when the predictors were highly correlated, the proposed method, similar to the existing methods, tended to exhibit diminished performance in both variable selection and prediction. However, the relative performance among methods remained consistent with the independent case, and the proposed method demonstrated consistently competitive or superior performance across all scenarios.

\begin{figure}[htbp]
    \centering
    \includegraphics[width = \textwidth]{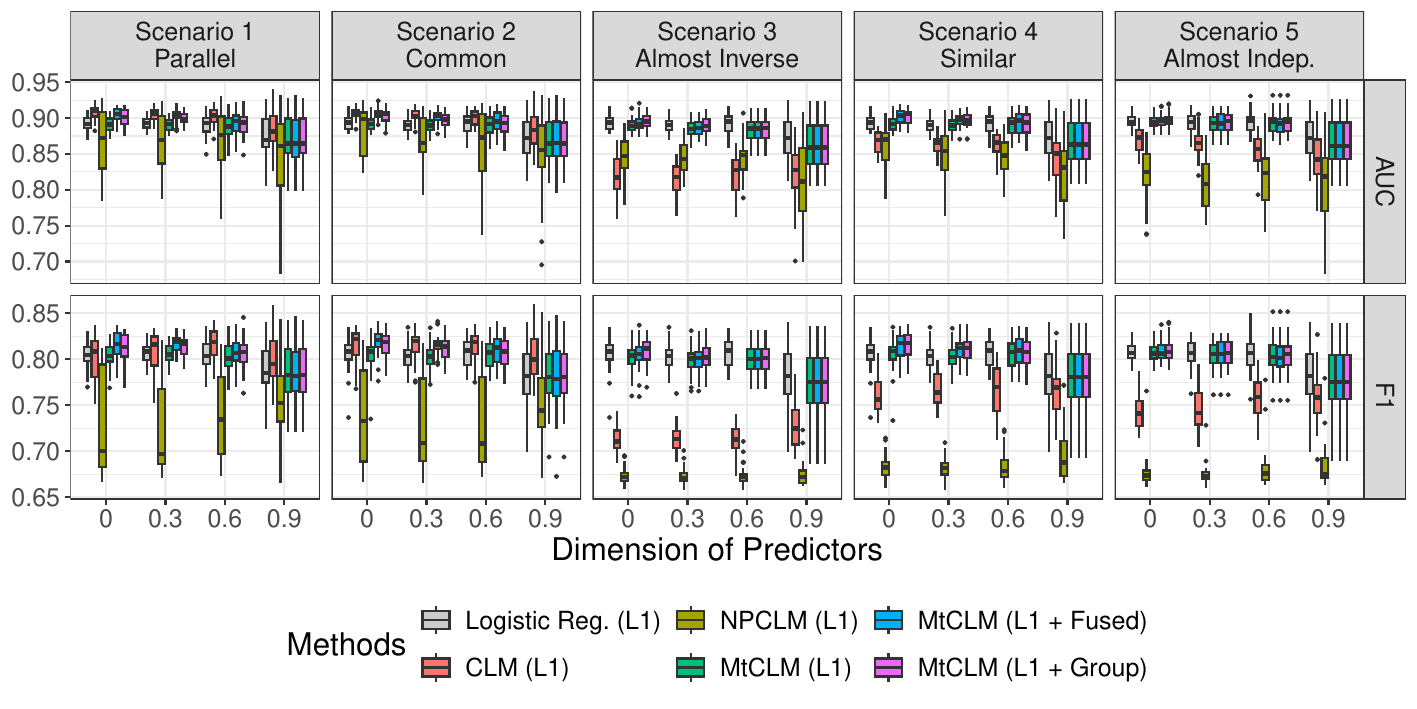}
    \caption{Comparison of the proposed and existing methods for the screening (0/1 classification) with 75-dimensional {\bf correlated} predictors.}
    \label{fig:performance_corr_screening_75}
\end{figure}

\begin{figure}[htbp]
    \centering
    \includegraphics[width = \textwidth]{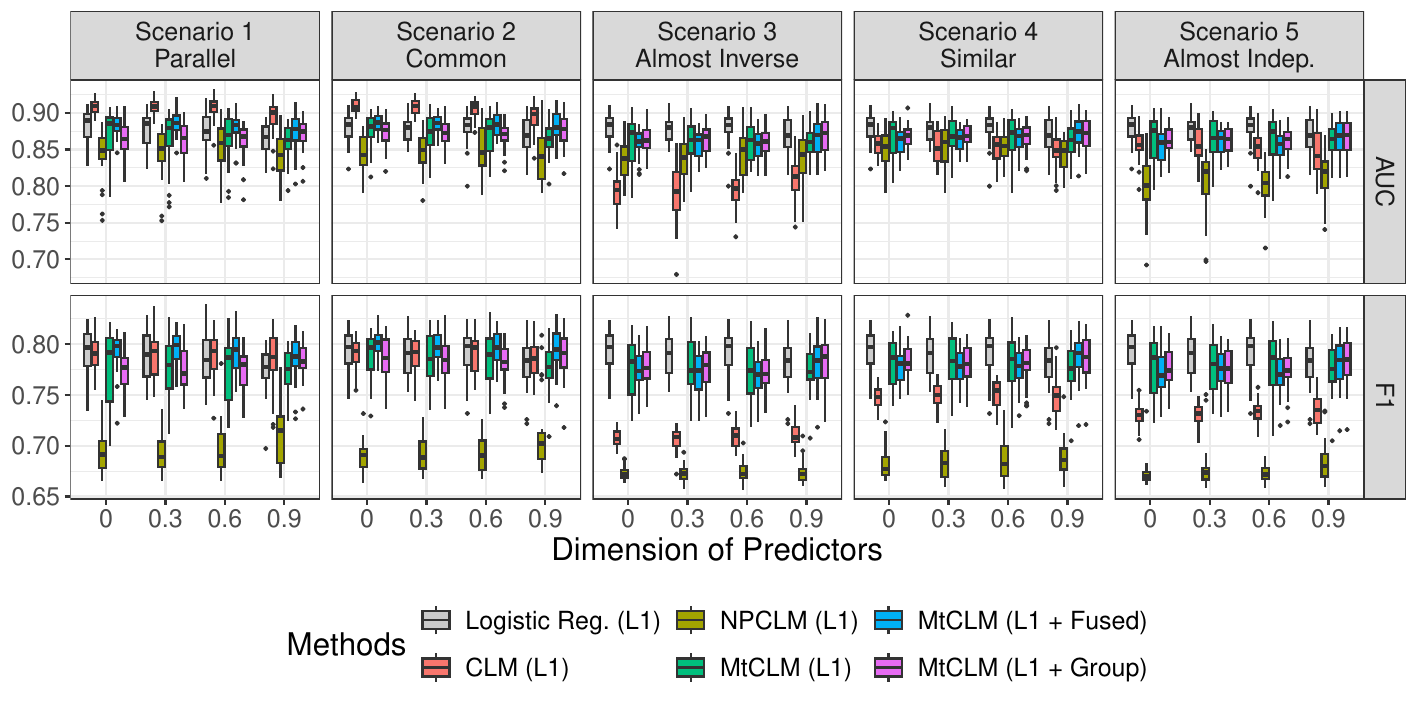}
    \caption{Comparison of the proposed and existing methods for the screening (0/1 classification) with 450-dimensional {\bf correlated} predictors.}
    \label{fig:performance_corr_screening_450}
\end{figure}

\paragraph{Screening Performance}
Figures \ref{fig:performance_corr_screening_75} and \ref{fig:performance_corr_screening_450} show the impact of correlation on screening performance in the low-dimensional case ($p = 75$) and the high-dimensional case ($p = 450$), respectively. The screening performance tended to deteriorate as the value of the Toeplitz correlation parameter $\rho$ increased. The relative performance between the proposed and existing methods showed a similar trend to that observed in the uncorrelated case.

\begin{figure}[htbp]
    \centering
    \includegraphics[width = \textwidth]{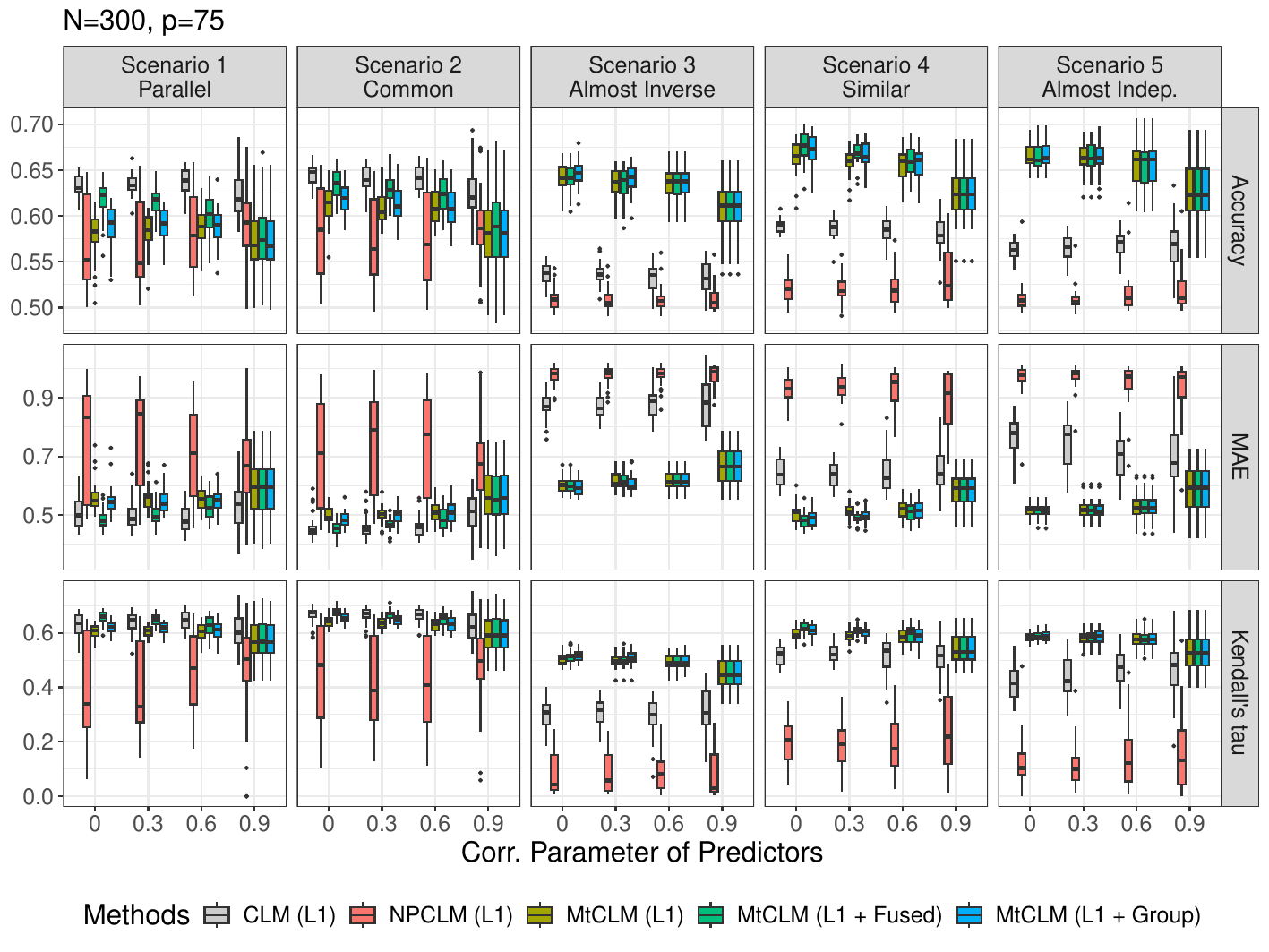}
    \caption{Comparison of the proposed and existing methods for the joint task of screening and severity prediction with 75-dimensional {\bf correlated} predictors.}
    \label{fig:performance_corr_sevpred_75}
\end{figure}

\begin{figure}[htbp]
    \centering
    \includegraphics[width = \textwidth]{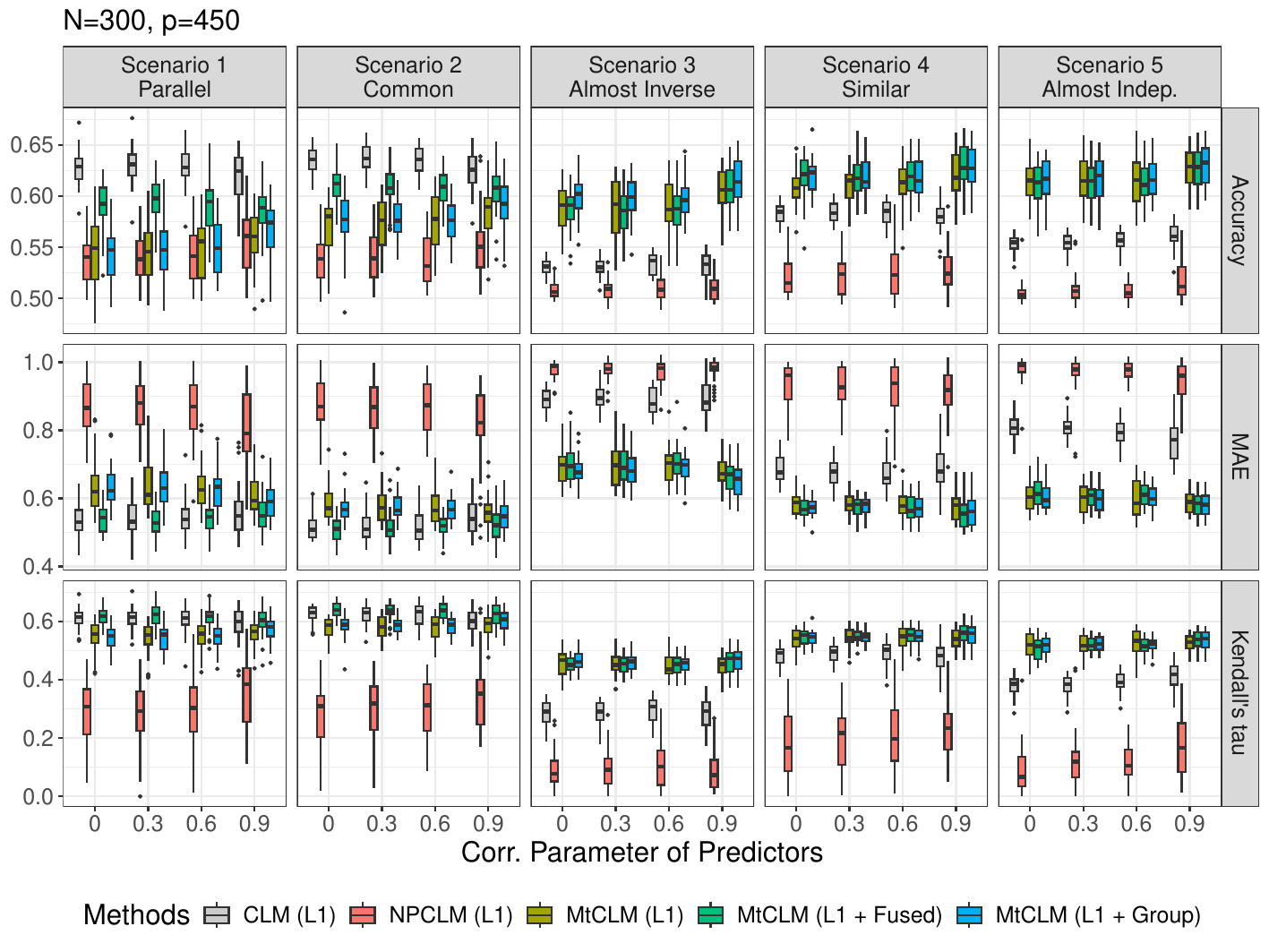}
    \caption{Comparison of the proposed and existing methods for the joint task of screening and severity prediction with 450-dimensional {\bf correlated} predictors.}
    \label{fig:performance_corr_sevpred_450}
\end{figure}

\paragraph{Severity Prediction Performance}
Figures \ref{fig:performance_corr_sevpred_75} and \ref{fig:performance_corr_sevpred_450} show the impact of correlation on the performance of the joint task of screening and severity prediction under the low-dimensional case ($p = 75$) and the high-dimensional case ($p = 450$), respectively. Similarly to the screening, the prediction performance tended to deteriorate as the value of the Toeplitz correlation parameter $\rho$ increased. The relative performance between the proposed and existing methods showed a similar trend to that observed in the uncorrelated case, but the advantage of the proposed method may be reduced when the correlation is as strong as $\rho = 0.9$.

\paragraph{Variable Selection Performance}
Figures \ref{fig:performance_corr_sevpred_75} and \ref{fig:performance_corr_sevpred_450} show the impact of correlation on the performance in variable selection under the low-dimensional case ($p = 75$) and the high-dimensional case ($p = 450$), respectively. 

As can be inferred from general properties of regression analysis beyond MtCLM, both FDR and power significantly deteriorated when strong correlations existed among predictors. In particular, when the correlation was as high as $\rho = 0.9$, the differences due to the structure of the penalty terms nearly disappeared, and all methods showed degraded performance in variable selection.

\begin{figure}[htbp]
    \centering
    \includegraphics[width = \textwidth]{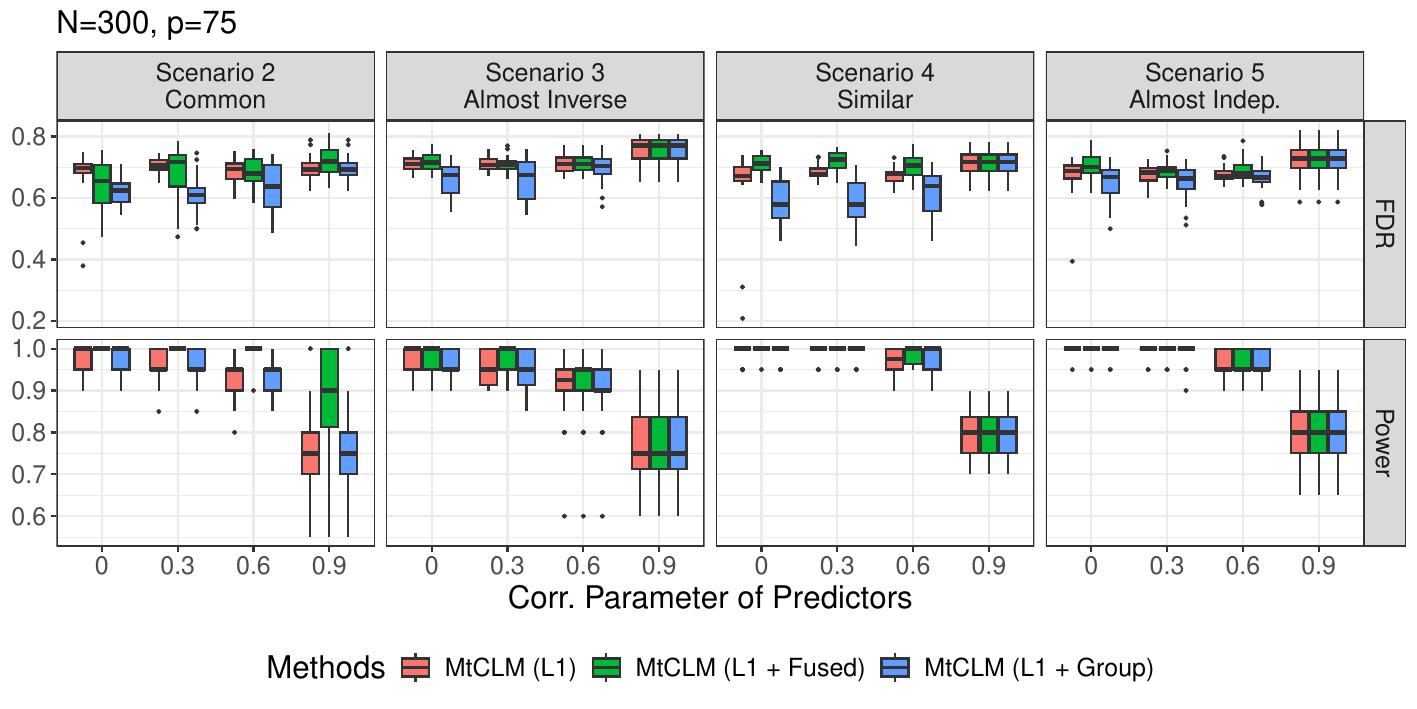}
    \caption{Comparison of the proposed methods in variable selection among 75-dimensional {\bf correlated} predictors.}
    \label{fig:performance_corr_selection_75}
\end{figure}

\begin{figure}[htbp]
    \centering
    \includegraphics[width = \textwidth]{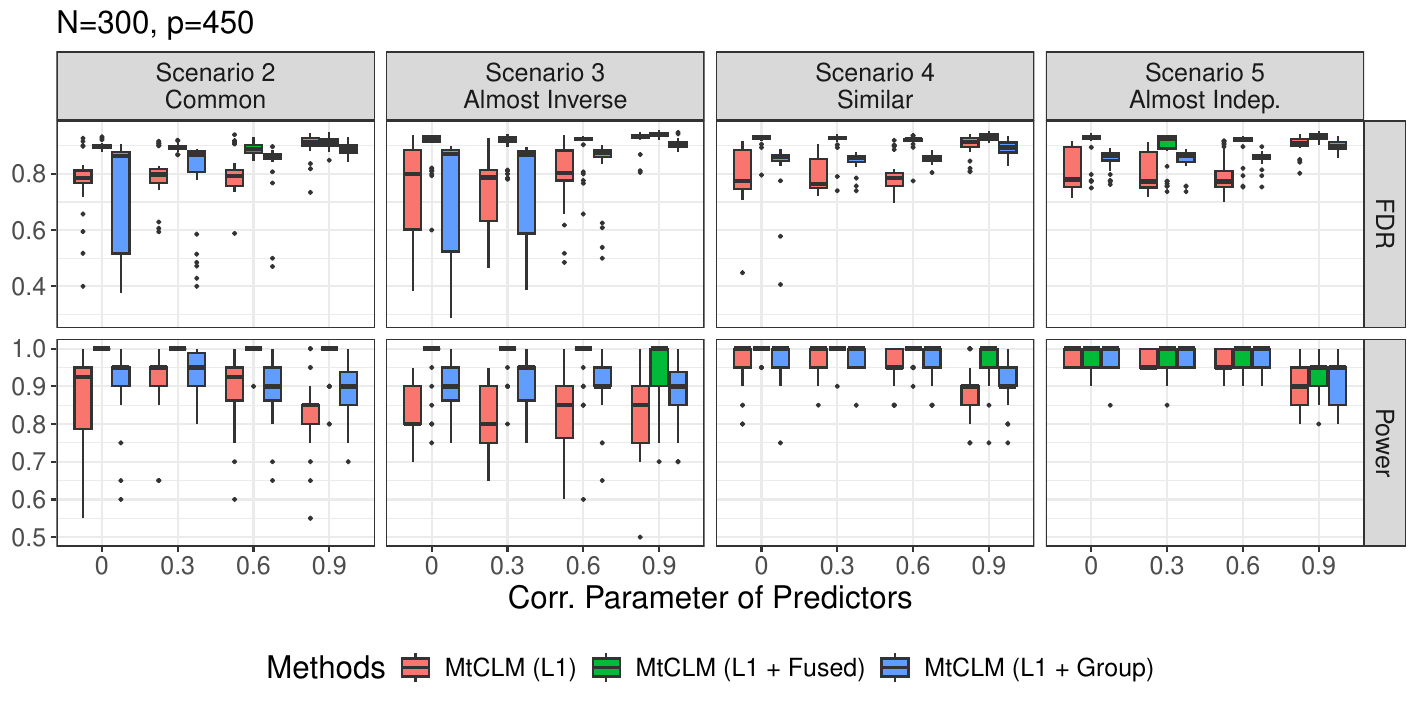}
    \caption{Comparison of the proposed methods in variable selection among 450-dimensional {\bf correlated} predictors.}
    \label{fig:performance_corr_selection_450}
\end{figure}

\clearpage
\section{Further Information of Real Data Analysis}\label{app:realdata}
\subsection{Pancreatic Ductal Adenocarcinoma Dataset}\label{app:PDAC}
\begin{figure}[ht]
    \centering
    \includegraphics[width = 0.9\textwidth]{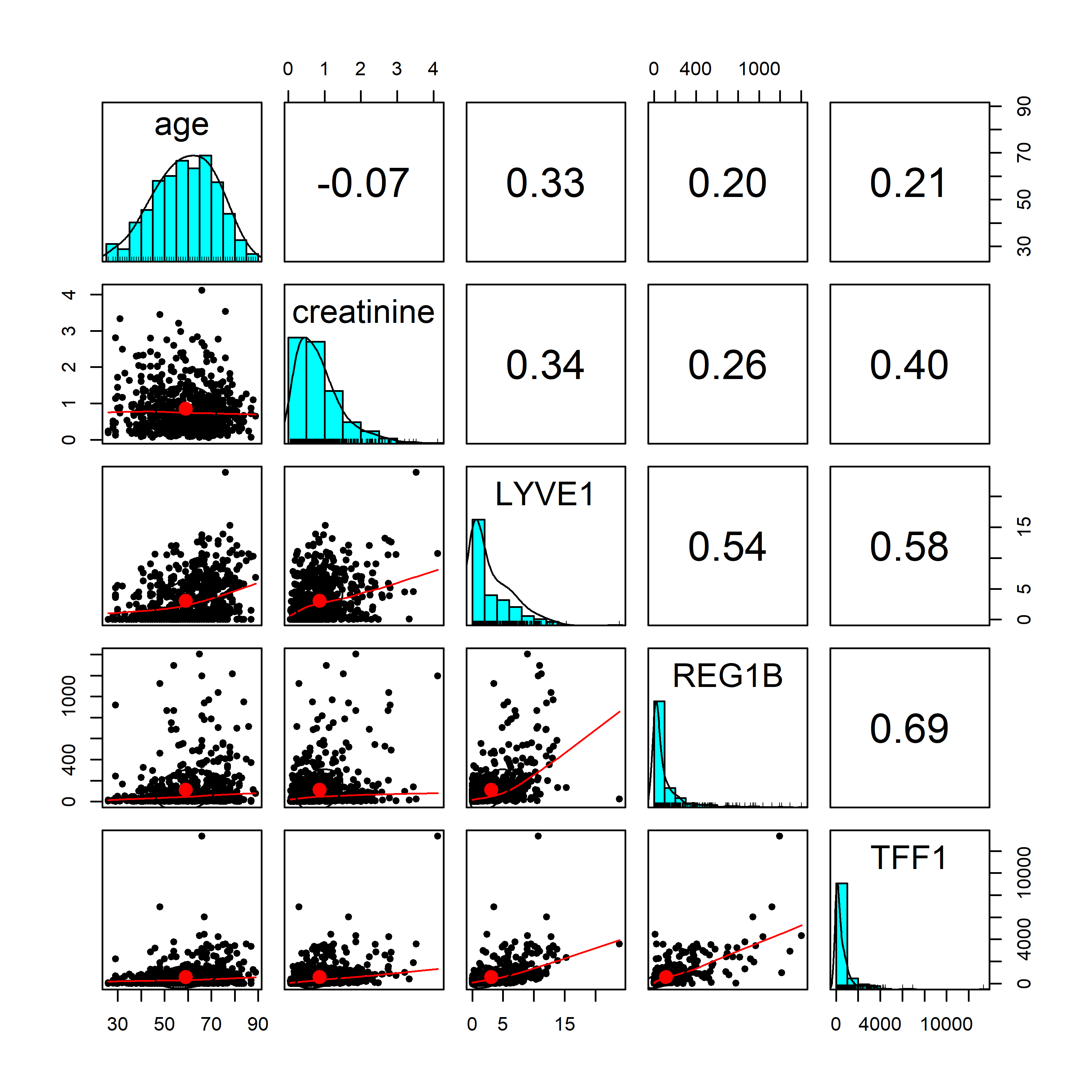}
    \caption{Correlation plot of age and biomarkers in the dataset provided by Debernardi et al. (2020). Each value in the upper-right boxes is Pearson's correlation.}
    \label{fig:Debernardi_corr}
\end{figure}

\begin{figure}[ht]
    \centering
    \includegraphics[width = 0.9\textwidth]{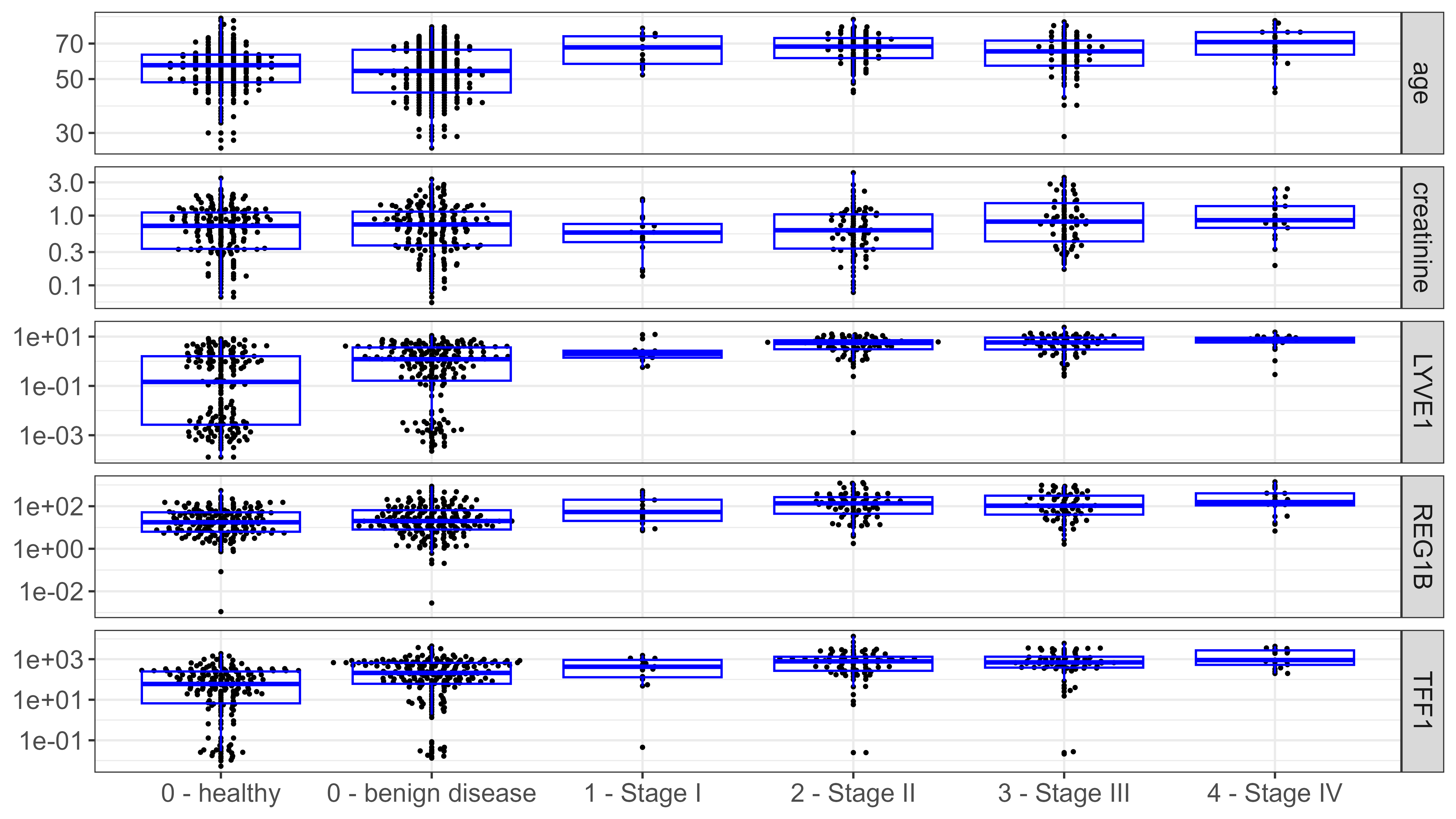}
    \caption{A box plot overlaid by beeswarm. This shows univariate relationships between markers and severity in the dataset provided by Debernardi et al. (2020).}
    \label{fig:Debernardi_univ}
\end{figure}

\clearpage
\subsection{METABRIC Cohort Dataset}\label{app:METABRIC}
\begin{figure}[htbp]
    \centering
    \includegraphics[width = 0.7\textwidth]{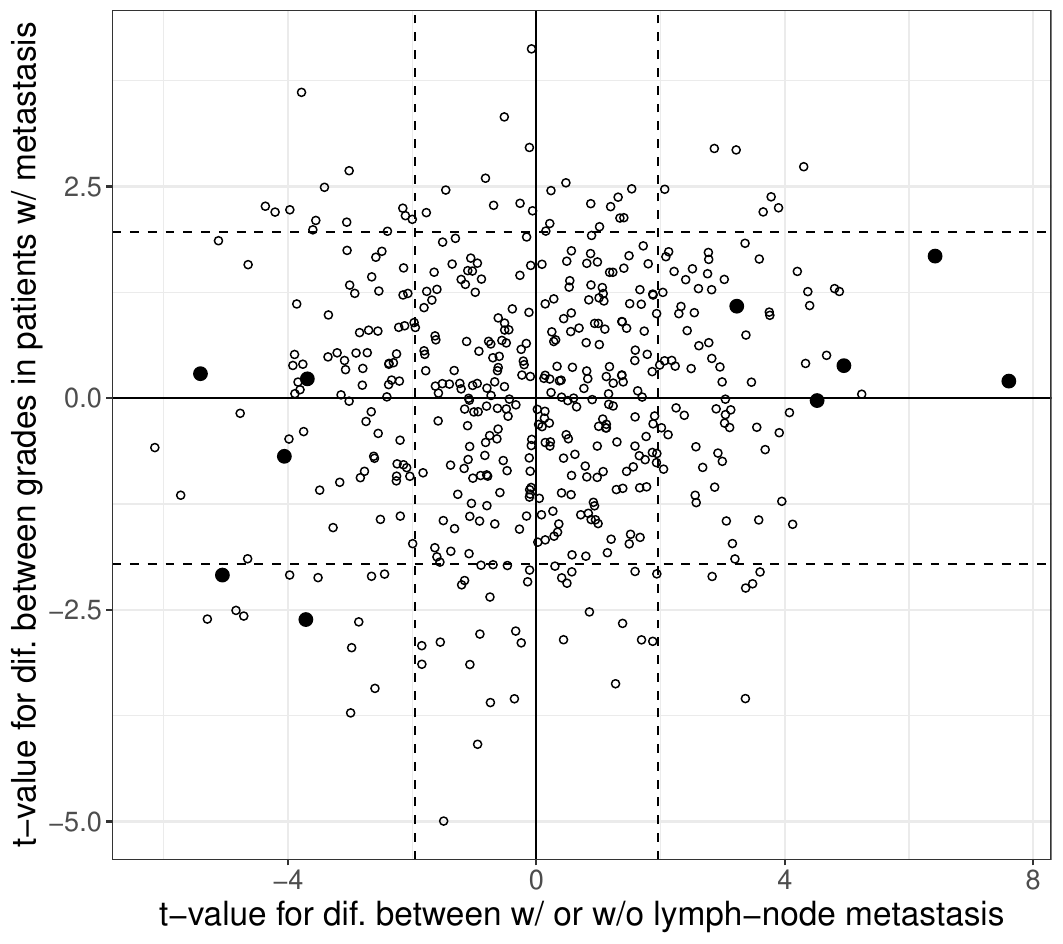}
    \caption{The relationship between the response and each mRNA level, and the selection by the L1-penalized logistic regression. The horizontal axis compares each mRNA level with the presence or absence of lymph node metastasis, and the vertical axis compares mRNA levels between Stage 2 and higher stages for the cases with lymph node metastasis. The filled dots represent predictors selected by the L1-penalized logistic regression.}
\end{figure}

\begin{figure}[htbp]
    \centering
    \includegraphics[width = 0.7\textwidth]{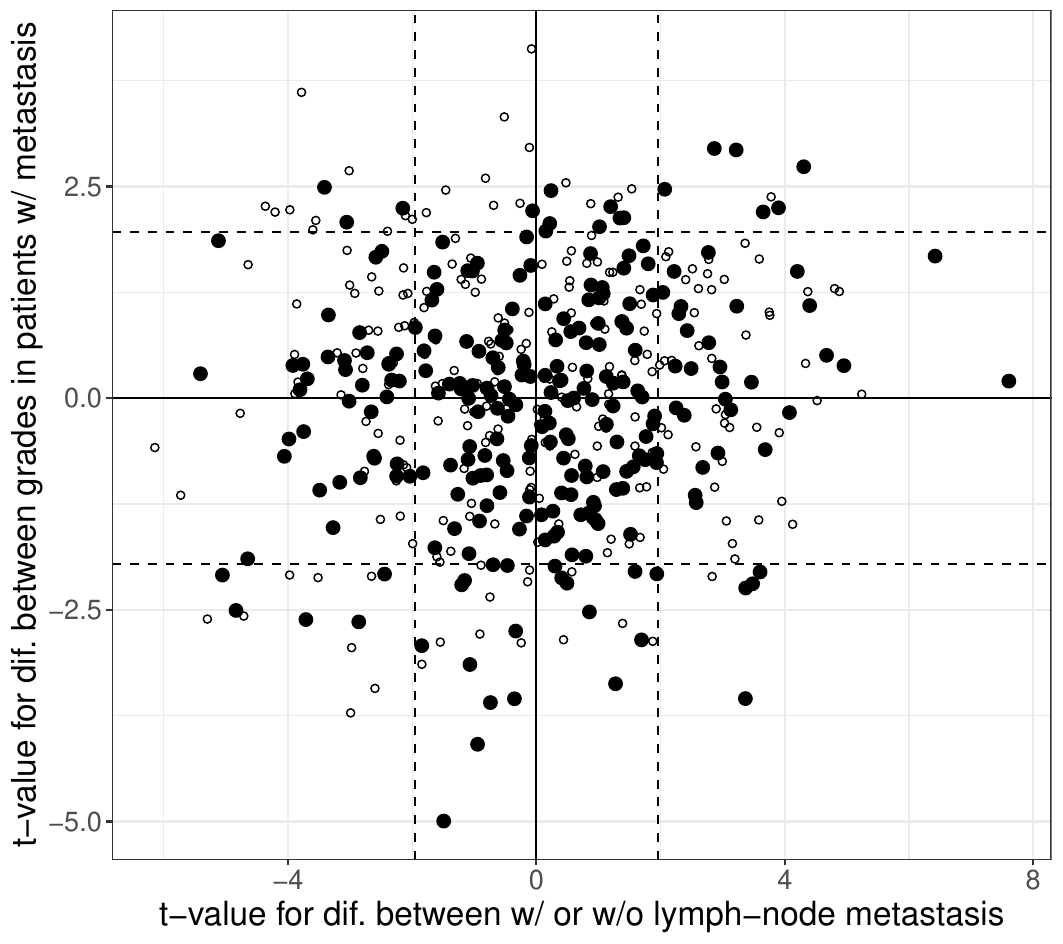}
    \caption{The relationship between the response and each mRNA level, and the selection by the L1-penalized parallel CLM. The horizontal axis compares each mRNA level with the presence or absence of lymph node metastasis, and the vertical axis compares mRNA levels between Stage 2 and higher stages for the cases with lymph node metastasis. The filled dots represent predictors selected by the L1-penalized parallel CLM.}
\end{figure}

\begin{figure}[htbp]
    \centering
    \includegraphics[width = 0.7\textwidth]{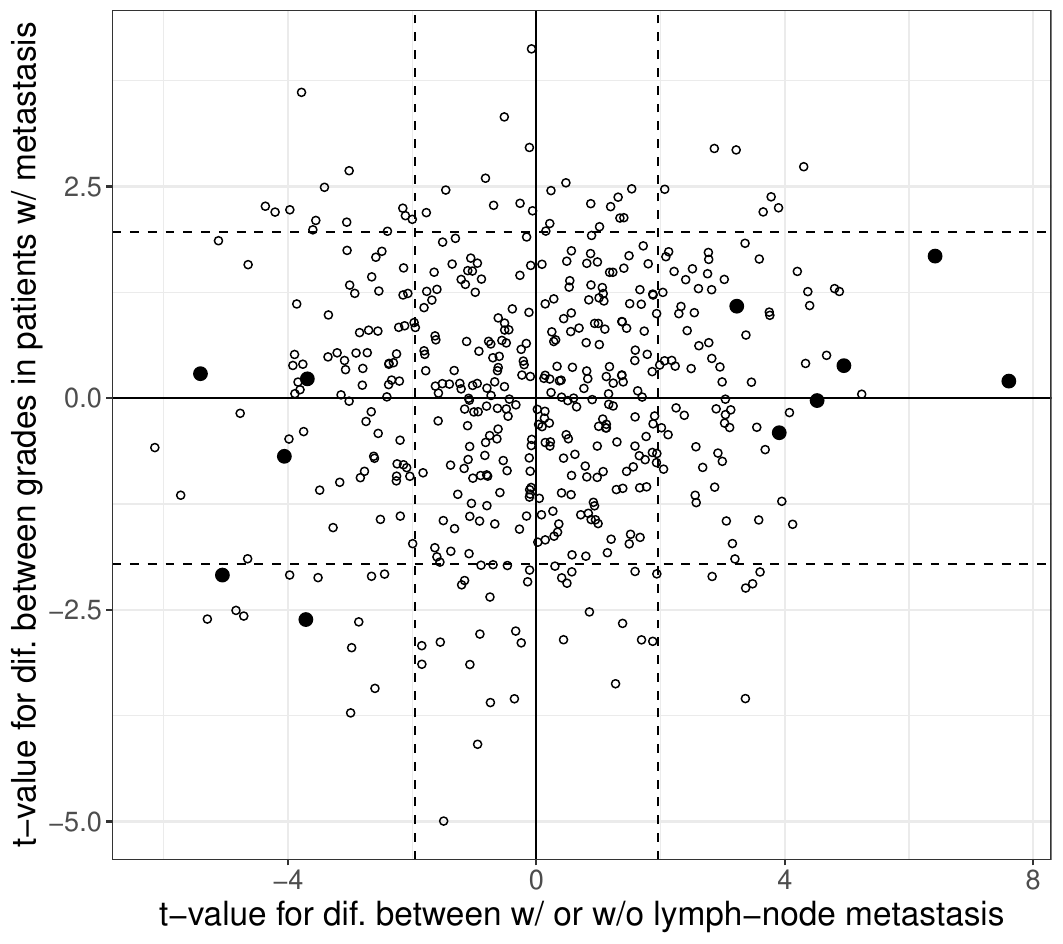}
    \caption{The relationship between the response and each mRNA level, and the selection by MtCLM (L1). The horizontal axis compares each mRNA level with the presence or absence of lymph node metastasis, and the vertical axis compares mRNA levels between Stage 2 and higher stages for the cases with lymph node metastasis. The filled dots represent predictors selected by MtCLM (L1).}
\end{figure}

\begin{figure}[htbp]
    \centering
    \includegraphics[width = 0.7\textwidth]{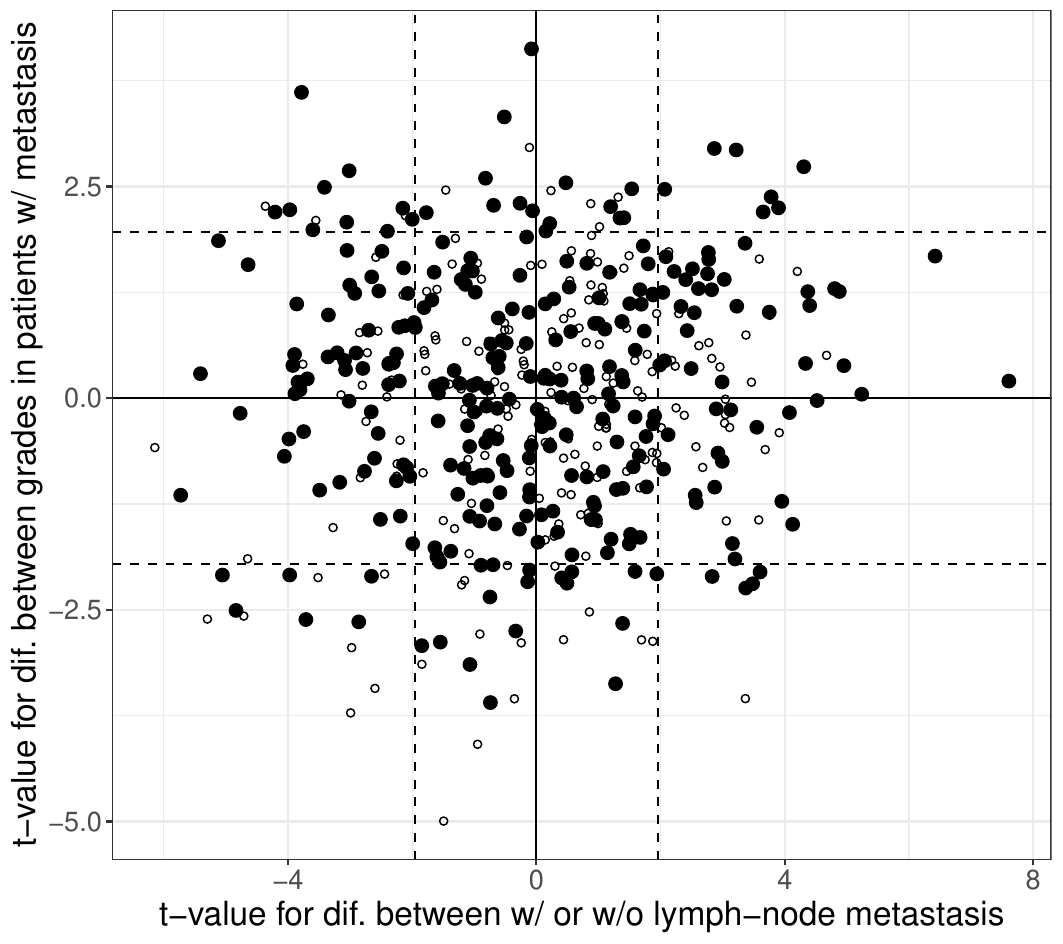}
    \caption{The relationship between the response and each mRNA level, and the selection by MtCLM (L1 + Fused). The horizontal axis compares each mRNA level with the presence or absence of lymph node metastasis, and the vertical axis compares mRNA levels between Stage 2 and higher stages for the cases with lymph node metastasis. The filled dots represent predictors selected by MtCLM (L1 + Fused).}
\end{figure}

\begin{figure}[htbp]
    \centering
    \includegraphics[width = 0.7\textwidth]{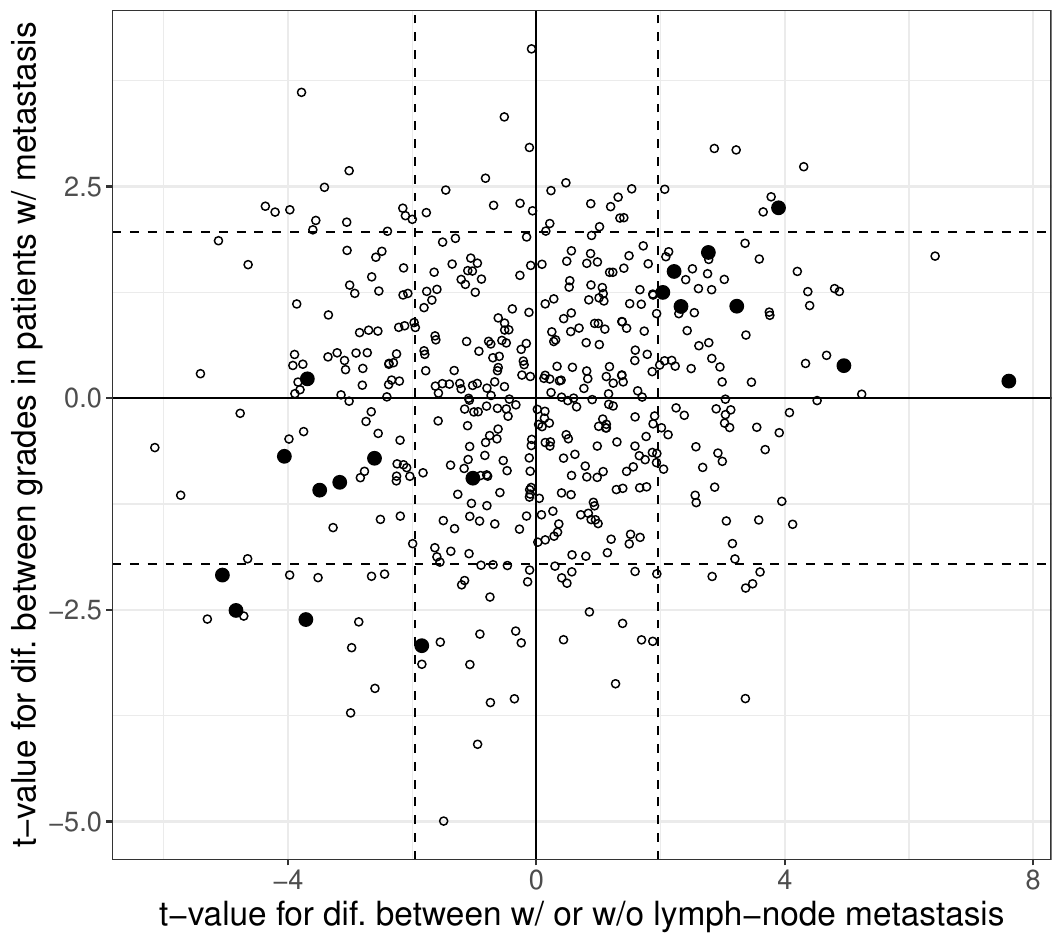}
    \caption{The relationship between the response and each mRNA level, and the selection by MtCLM (L1 + Fused) with thresholding. The horizontal axis compares each mRNA level with the presence or absence of lymph node metastasis, and the vertical axis compares mRNA levels between Stage 2 and higher stages for the cases with lymph node metastasis. The filled dots represent predictors selected by MtCLM (L1 + Fused), whose absolute values of the regression coefficients were larger than 0.01.}
\end{figure}

\clearpage
{\colr
\begin{landscape}
\begin{table}
    \centering
    \caption{Variables associated with breast cancer metastasis or grade that were identified by the proposed and comparative methods. The rightmost column lists example references that discuss the association between each gene and breast cancer.}\vspace{2mm}
    \begin{tabular}{llccccccl} \hline \label{tab:appMETABRIC}
         &  & Logistic & CLM & \multicolumn{4}{c}{MtCLM} & \\\cline{5-8}
        Type & Gene & L1 & L1 & L1 & L1+Fused & L1+Fused+Cutoff & L1+Group &  Example References  \\ \hline
        Expression & {\it BARD1} & \checkmark &  & \checkmark & \checkmark &  & \checkmark & \citet{Wu1996-pa} \\
         & {\it STAT5B} & \checkmark & \checkmark & \checkmark & \checkmark & \checkmark & \checkmark & \citet{Peck2011-sf}  \\
         & {\it RBPJ} & \checkmark & \checkmark & \checkmark & \checkmark & \checkmark & \checkmark &  \citet{Shi2022-qh} \\ 
         & {\it AURKA} & \checkmark & \checkmark & \checkmark & \checkmark &  & \checkmark & \citet{Wang2006-au}  \\ 
         & {\it CASP10} & \checkmark & \checkmark & \checkmark & \checkmark & \checkmark & \checkmark & \citet{Frank2006-ed}  \\
         & {\it DIRAS3} & \checkmark & \checkmark & \checkmark & \checkmark & \checkmark & \checkmark & \citet{Yu1999-cn}  \\
         & {\it GSK3B} & \checkmark & \checkmark & \checkmark & \checkmark & \checkmark & \checkmark & \citet{Quintayo2012-te} \\
         & {\it RPS6KA2} & \checkmark & \checkmark & \checkmark & \checkmark &  &  & \citet{Serra2013-gc} \\ 
         & {\it SMAD2} &  &  & \checkmark &  &  &  & \citet{Samanta2012-rv} \\ 
         & {\it RUNX1} & \checkmark & \checkmark & \checkmark & \checkmark & \checkmark & \checkmark & \citet{Chimge2016-kh} \\
         & {\it HSD3B1} & \checkmark & \checkmark & \checkmark & \checkmark & \checkmark & \checkmark & \citet{Kruse2021-vn} \\
        Mutation & {\it FANCD2} & \checkmark & \checkmark & \checkmark & \checkmark & \checkmark & \checkmark & \citet{Mantere2017-gl}  \\\hline
        \multicolumn{2}{l}{\# Selected (/581)}  & 11 & 314 & 12 & 324 & 21 & 10 &   \\ \hline
    \end{tabular}
\end{table}
\end{landscape}
}

\end{document}